\documentclass[a4paper,11pt]{article}
\pdfoutput=1
\usepackage[utf8]{inputenc}
\usepackage[T1]{fontenc}
\usepackage{lmodern}
\usepackage[british]{babel}
\usepackage{amsmath,slashed}
\usepackage{amssymb}
\usepackage{amsfonts}
\usepackage{float,physics}
\usepackage{stmaryrd}
\usepackage{url}

\usepackage{graphicx,psfrag}
\usepackage{placeins,bbm}
\usepackage{subcaption}

\usepackage{jheppub}

\DeclareOldFontCommand{\bf}{\normalfont\bfseries}{\mathbf}
\newcommand{\ZZ}{\ensuremath{\mathbb{Z}}}

\DeclareMathOperator{\Pf}{Pf}

\newcommand{\Dsla}{{\mbox{\, $\!\!\not\!\!D$}}}
\DeclareMathOperator{\R}{\!R}

\newcommand{\Oopm}{O^{\pm}}
\newcommand{\Oop}{O^{+}}
\newcommand{\Oom}{O^{-}}

	\title{Identifying topological  structures with adjoint mode filtering}

    \author[c]{Georg Bergner}
    \author[a,b]{Antonio Gonz\'alez-Arroyo }
    \author[c]{Ivan Soler}
    
    \affiliation[a]{ Instituto de F\'{\i}sica Te\'orica UAM/CSIC,  Nicol\'as
      Cabrera 13-15, \\
      Universidad Aut\'onoma de Madrid, E-28049 Madrid, Spain}
      
    \affiliation[b]{
     Departamento de F\'{\i}sica Te\'orica,  M\'odulo 15,
     \\
      Universidad Aut\'onoma de Madrid, Cantoblanco, E-28049 Madrid, Spain}
      
    \affiliation[c]{University of Jena, Institute for Theoretical Physics, \\
     Max-Wien-Platz 1, D-07743 Jena, Germany}

    \emailAdd{antonio.gonzalez-arroyo@uam.es}
    \emailAdd{georg.bergner@uni-jena.de}
    \emailAdd{ivan.soler.calero@uni-jena.de}

    \abstract{We present new investigations on the Adjoint Filtering Method (AFM), a proposal for filtering gauge configurations by using adjoint zero modes. This method relies on the existence of the Supersymmetric Zero Mode (SZM), whose density follows the gauge action density for classical configurations. We review how a similar construction on the lattice is implemented based on the overlap operator and test the method against smooth configurations showing a remarkable agreement with the expected densities even when pairs of fractionals instanton/anti-instantons are present and noise is added. Then we explore the application of the method to Monte Carlo generated configurations based on SU(2) gauge group. The tuning of the parameters and quantitative results are explicitly shown for a $T^3\times R$ lattice. We show explicit examples comparing the AFM to the density obtained from the Gradient Flow. The agreement is remarkable for some specific configurations containing fractional instantons with the advantage that the AFM does not modify the underlying structures.}
     
\begin{document}
	\maketitle
\section{Introduction and motivation}
Lattice gauge theories have proven to be a very useful tool to
investigate non-perturbative properties of gauge theories. This includes the phenomenologically interesting case of QCD,  allowing
a first-principles computation of many experimental quantities
involving strong interactions.\footnote{Latest results are summarized in the FLAG report~\cite{FlavourLatticeAveragingGroupFLAG:2021npn}.} In addition, there are other field
theories which are of great interest either for being candidate
extensions of the Standard Model or for their theoretical interest.
The latter includes the case of Yang-Mills  theory and its
supersymmetric extensions. The standard computational techniques used
in Lattice Gauge Theory studies are dominated by the Monte Carlo
method and related importance-sampling numerical techniques. There is
however a challenge  in trying to achieve a qualitative and
semiquantitative understanding of the origin and magnitude of the
non-perturbative phenomena involved. 
Our approach to describing strongly interacting theories is based on a 
semiclassical analysis, which has been demonstrated to work at least in certain regimes, as explained towards the end of this Section.
Apart from searching for new formulations 
to describe strongly interacting phenomena and allow analytical computation,
 it is clear that analysing
samples of lattice configurations can provide answers or clues 
regarding the underlying origin of these phenomena as well as new analytical methods.
This is the idea which serves as a motivation for the present work. 

Lattice configurations describe snapshots of the vacuum of the theory and 
this includes its structure at different length scales. It is to be
expected that the most relevant length scale is given by the inverse mass gap
or inverse Lambda parameter of the theory. Nonetheless, 
quantum fluctuations grow at short length scales as a result of the
ultraviolet divergences of these theories. In practice, this implies that the
action density, for example,  is dominated by this short-wavelength fluctuations and it is hard to see any underlying long range structure
appearing at the physical scale mentioned earlier. Extended semiclassical objects like
instantons and fractional instantons are expected to be present but
they are completely covered up by the noise. In this paper we are
going to explore a filtering method that
can eliminate the noise and allow to uncover these features. The
ultimate goal would be to unveil secrets contained in these configurations. 

The general idea has a longer history. A special role was initially
played by different iterative smoothening techniques such as cooling or smearing~\cite{Teper:1987wt,
Teper:1987ws, Ape:1987thf, Fernandez:1987ph, Teper:1985rb}. 
These methods are local and have a characteristic smoothening radius that has to be
monitored carefully in order to erase the high
frequency fluctuations without altering the contents at the relevant 
scales one wants to explore.
Disregarding this careful analysis has sometimes led to unjustified general critiques of the method.
The panorama has been greatly clarified
with the introduction of gradient flow techniques in which the
smearing radius becomes a continuous parameter~\cite{Lohmayer:2011si, Luscher:2010iy,Narayanan:2006rf}. 
The effects are quite similar to other smoothening techniques, but the method has a well-defined continuum limit~\cite{Luscher:2010iy}. More recently the effects of the 
gradient flow on instanton properties have been monitored and even an extrapolation in the flow time has been applied~\cite{Athenodorou:2018jwu}.
Despite the fact that they have proven to be very useful in the analysis of configurations, smoothening methods have also well studied limitations.
Smoothening provides a limited window for the underlying structures of the gauge configurations altering some structures outside of it as we explained in later sections of this work.

The criticism on cooling techniques has been an initial motivation to study alternative approaches. 
More generally, the limitations of the smoothening make an alternative point of view, which does not so much rely on the change of the gauge field, desirable.
More importantly, the semiclassical analysis aims for a complete description of the relevant features of the QCD vacuum. Therefore one should include alternative observables and approaches to confirm that a consistent picture emerges.
To demonstrate the presence of instanton structures in the configurations, several
authors~\cite{Gattringer:2002gn,Bruckmann:2006wf,Solbrig:2007nr} made use of the solution of the Dirac equation in 
the gauge field background. In the continuum, solutions of the gauge field equations of motion like the BPST instanton ~\cite{Belavin:1975fg}
or the KvBLL calorons~\cite{Lee:1998vu, Lee:1998bb,Kraan:1998sn,Kraan:1998pm} have associated solutions of the
Dirac equation in the fundamental ~\cite{tHooft:1976snw,Grossman:1977mm, Corrigan:1978ce,GarciaPerez:1999ux,Chernodub:1999wg} and adjoint representation~\cite{GarciaPerez:2006rt,GarciaPerez:2008gw, GarciaPerez:2009mg}, whose densities trace back the
presence of these structures. The advantage is that under the presence
of high frequency quantum noise, these densities are less affected
than the corresponding gauge field observables. The  situation becomes
optimal if we consider the Dirac equation in the adjoint
representation because in that case one particular solution of the
Dirac equation (the supersymmetric zero-mode SZM) has a density which 
is identical to the action density of the gauge field. When adding
noise to the gauge field, the corresponding  SZM density implements a
filtering of this noise with respect to the action density itself.
This idea was the germ of a proposal made a few years ago by one of the authors 
and collaborators~\cite{Gonzalez-Arroyo:2005fzm, GarciaPerez:2011tx}
to produce an effective
filtering method applicable to lattice configurations. In those papers
a special property of the SZM was found which distinguishes it  from
other solutions. In those papers tests were made to show that the
filtering was working properly for some smooth configurations before
and after applying noise to them. However, the ultimate purpose of
applying this methodology to the relevant Monte Carlo generated
configurations was not attempted and in this publication we will try
to take the first steps in this direction. Note that the adjoint representation has been considered
also in different context for a numerical analysis of gauge field configurations~\cite{Hollwieser:2010mj,Hollwieser:2012kb}.

What features of the QCD or pure Yang-Mills vacuum are we aiming to identify?  Our
goal is  to explore  the validity of semiclassical ideas. 
This has a long history starting with the work of
Callan, Dashen, and Gross~\cite{Callan:1977qs, Callan:1977gz} based on merons. Other ideas in
the same spirit are the instanton liquid model of Shuryak~\cite{Shuryak:1981ff},
and the fractional instanton liquid model (FILM) proposed by one of
the present authors~\cite{GonzalezArroyo:1995zy,GonzalezArroyo:1995ex}. Our belief in semiclassical ideas is
supported by theories in lower dimensions~\cite{Polyakov:1975rs} and with additional matter
content~\cite{Seiberg:1994rs,Seiberg:1994bz,Unsal:2007jx}. 
In this program the analysis of lattice configurations 
is expected to play a crucial role. 
It is essential, in our opinion, to consider first regimes where the semiclassical analysis is under control. 
In such kind of regimes, the numerical method can be tested reliably. Moreover, such a controlled regimes can be even continuously connected to full vacuum and hence it is possible to conjecture 
more fundamental insights from such an analysis.
Monitoring this process on the lattice one expects to find clues to validate these ideas. 

An specific regime in which the semiclassical ideas and methodology is expected to work is that in which the theory is formulated on a compact manifold in which the size in some directions is small. The validity of semiclassical analysis in this regime is a consequence of asymptotic freedom. This has already
been proven by the studies carried out many years ago by one of the
authors and collaborators~\cite{GarciaPerez:1993ab}. The studies were
performed for a lattice of size of size $L_s^3 \times L_0$, in which
$L_s \ll L_0$. It corresponds to a Hamiltonian study of
Yang-Mills theory in a small spatial box. This was inserted into a
program initiated by L\"uscher~\cite{Luscher:1982uv} of studying the transition
from small to large volumes in asymptotically free theories. The
results showed clearly how  perturbative and semiclassical methods were
successful in describing the lattice results in a range of physical
spatial sizes. A leading role in this analysis was played by
fractional instanton solutions which tunnel between perturbative vacua
and lead to a recovery of isotropy and confinement~\cite{GonzalezArroyo:1995zy, GonzalezArroyo:1995ex, GonzalezArroyo:1996jp}.
It is on the basis of these results that the authors proposed a
picture of the Yang-Mills vacuum based upon a liquid of fractional
instantons which could explain and relate a finite topological susceptibility and
string tension. Recently there has been a renewed interest in fractional instantons and their dynamical role both in  analytic work~\cite{GarciaPerez:2000aiw, Gonzalez-Arroyo:2019wpu, Unsal:2020yeh,  Anber:2022qsz,Poppitz:2022rxv,Anber:2023sjn,Nair:2022yqi}
and in numerical one~\cite{Itou:2018wkm, Mickley:2023exg}. 
In most of the afore-mentioned studies at finite volume an important role
is played by boundary conditions: the twisted boundary conditions (TBC)
introduced by `t Hooft~\cite{tHooft:1979rtg, tHooft:1980kjq, tHooft:1981sps}. Indeed, boundary conditions are
crucial at small volumes but should become irrelevant at large ones. 
Semiclassical arguments suggest why TBC are more effective in
revealing the transition from small to large volumes. Indeed, even
in the absence of twist,  solutions of the  Yang-Mills classical
equations of motion appear that are made out of assemblies of
fractional instantons. Twist only imposes a constraint on the number
of them. This will become clear with the configuration examples
studied in this paper.  

Apart from the previous Hamiltonian picture with three small directions, that we can describe as
a $T^3\times R$ topology, more general topologies can be considered. 
If two directions ($L_1$ and $L_2$) are much smaller that the other two ($L_0$ and $L_3$), a $T^2\times R^2$ topology is obtained.
Again we consider twisted boundary conditions in this setup and fractional instanton structures also play a role here. 
These solutions were studied and described as 
vortex-like~\cite{GonzalezArroyo:1998ez,
Montero:1999by, Montero:2000pb}, since they can be  looked upon as $R^4$ solutions
periodic in two directions, which act as the world sheet directions in
this setting.  This two-dimensional picture is particularly attractive
as the corresponding fractional instantons are well separated since
their action-density profile is exponentially localized. This allows
creating configurations containing  both fractional instantons and
anti-instantons. Although these configurations are not exact solutions of the classical equations of motion, when  the separation of structures is large they give rise to quasi-zero modes of the Dirac operator that
can be neatly identified. Furthermore, despite not being protected by
a non-zero topological charge they can be extremely stable under
gradient flow. The vortex-like fractional instantons do carry center
flux as shown in the original papers~\cite{GonzalezArroyo:1998ez,
Montero:1999by, Montero:2000pb}. They can provide
an underlying gauge-invariant substrate~\cite{Montero:1999by} to the center-vortices obtained after gauge-fixing and studied by many lattice
studies~\cite{DelDebbio:1997ke, Engelhardt:1998wu,deForcrand:1999our, Greensite:2016pfc, Kamleh:2023gho}. 
The fact that there is a regime in which
they can be studied by semiclassical methods is very attractive and
connects with recent studies by other authors which also considered the 
$T^2\times R^2$ regime~\cite{Tanizaki:2022ngt, Anber:2022qsz}. 

The third geometric situation $R^3\times S^1$ is special in several
ways. It is the only situation in which a  $Q=1$ analytic solution was found showing  the
decomposition of the action into well-separated fractional instanton
structures: the KvBLL calorons~\cite{Lee:1998vu, Lee:1998bb,Kraan:1998sn,Kraan:1998pm}. 
In pure gauge theory, this geometry corresponds to finite temperature field theory, where the temperature
corresponds to the inverse of the $S^1$ radius. The semiclassical regime at small radius is therefore separated 
by the deconfinement transition from the physics of the confined vacuum.
However, if the theories are deformed, the transition can be avoided and a continuous connection between the semiclassical regime and the confined
vacuum is established; this is the basic approach behind different studies of adiabatic continuity~\cite{Unsal:2007jx, Anber:2018iof, Shifman:2008ja, Bergner:2018unx}. 
The most well established deformation is $\mathcal{N}=1$ supersymmetric Yang-Mills theory on $R^3\times S^1$ with periodic boundary conditions for the fermions. From the lattice perspective there have been a lot of work studying correlations  of the finite temperature behaviour of the  theory with topological structures (See for example ~\cite{Bornyakov:2015xao, 
Vig:2019wei, Larsen:2019sdi, Larsen_2022} and references therein).

Notice, that from the  fractional instanton perspective  all these
geometric situations are connected among themselves by Nahm
transform~\cite{Nahm:1979yw, Corrigan:1983sv,Schenk:1986xe, Braam:1988qk, donaldson1990geometry, GonzalezArroyo:1998ia} and form a unified picture~\cite{GarciaPerez:1999bc}.

In our opinion all of these investigations have shown in a very consistent way that
the semiclassical approach can indeed provide an understanding of the QCD vacuum. 
They still need to be extended and improved in several ways and input from numerical simulations is essential.
The final understanding of the underlying long-range structure of
lattice configurations is the most difficult and complex task and
remains a challenge. Whether the semiclassical regimes previously
mentioned help in disentangling this problem in the spirit of the
proposal made in Refs.~\cite{GonzalezArroyo:1995zy, GonzalezArroyo:1995ex} (see Ref.~\cite{Gonzalez-Arroyo:2023kqv} for a
recent summary) remains to be seen. Presumably, effort by many
researchers would be needed to achieve this goal. Nevertheless, as
this paper illustrates, the present study reveals rather beautiful structures 
appearing in  Yang-Mills theory. 

In the present work we improve the methods for the filtering required to study the long-range structures on Yang-Mills configurations. 
As explained, the main two categories are smoothening techniques and methods based on low modes of the Dirac-Wilson operator. We are comparing
two methods which are, for the reasons discussed above, the most advanced techniques in each of the two categories: the Gradient Flow (GF) and the Adjoint Filtering Method (AFM). 
These two techniques are explained in Sec.~\ref{sec:filtering_techniques}. In Sec.~\ref{sec:sclassical_conf} we introduce the semiclassical regime with
twisted boundary conditions and how we prepared test configurations with a given semiclassical background. Smooth test configurations as well as configurations with a controlled noise level 
are investigated in Sec.~\ref{smooth_configurations}. This confirms the correctness of our numerical methods and provides a first hint concerning the specific properties of the semiclassical contributions.
Finally, in Sec.~\ref{sec:mc_configs} Monte Carlo generated ensembles of configurations for pure Yang-Mills theory on $T^3\times R$ are analyzed. 
We have chosen the simulation parameters such that a controlled semiclassical behaviour is expected, but still ensuring that the density of topological excitations is large enough. In this way we can show how the 
optimization of the parameters for the methods can be achieved, which is in general quite challenging for a complete ensemble of configurations.

\section{Filtering techniques}
\label{sec:filtering_techniques}
As mentioned in the introduction, we will study different filtering techniques such that applied to Monte Carlo generated lattice configurations could resolve the underlying presence of topological structures that could explain the long-range features of the theory. 
All techniques have pros and cons and the use of them can show the robustness of the results and lead to an  optimal combination.

The quantities that are of interest for our study are the action density $S(x)$ and the topological charge density $q(x)$ related to long range seminclassical contributions. These can be defined from the gauge fields as
\begin{align}
\label{eq:SQField}
S(x)=\frac{1}{4g^2}F_{\mu \nu}^a(x)F_{\mu \nu}^a(x)\; , \quad 
q(x)=\frac{1}{32\pi^2} F_{\mu\nu}^a(x)
\tilde{F}_{\mu \nu}^a(x)\; .
\end{align}

The simplest  lattice implementation of these quantities comes from expressing the field strength components  in terms of clover plaquettes. This definition we denote as $q_c(x)$, $S_c(x)$ if not combined with smoothing techniques.

\subsection{Gradient flow}
\label{sec:GF}
As mentioned in the introduction this method is an improvement of the cooling/smearing methods that were used in the early studies. In the continuum the gradient flow evolves a given gauge field configuration from an initial value following the gradient of the Yang-Mills action functional. Thus, the action decreases with the flow parameter $\tau$. Intuitively, the effect is to smear out the field over an extension $\sqrt{8 \tau}$, the smearing radius. Thus, quantum fluctuations at scales smaller that this radius are strongly reduced without modifying substantially the long range structure present in the original configuration. The total topological charge of the configuration does not change under this continuous process. However, the flow process  modifies the initial configuration, for example by  erasing instanton/anti-instanton pairs separated by distance smaller that the smearing radius. This is not the only effect that the flow can have. If the base manifold is a torus, which is a particular relevant topology for lattice studies, there is a breaking of scale invariance leading to distortion of the size distribution of instantons. This is particularly dramatic for $Q=1$ since the limiting instanton size tends to zero. Indeed, for this reason there are no (self-dual) instantons on the torus as proven in Ref.~\cite{Braam:1988qk}.  

All this flow dynamics should be well understood before going over to the lattice version of the flow. On the lattice there are new challenges appearing. The choice of the lattice action used for the gradient is not irrelevant. Like the lattice size, also lattice artefacts destroy the 
scale-invariance of the theory. In practice, as studied in Ref.~\cite{GarciaPerez:1993lic}, this induces a modification of the size parameter of an instanton by the flow. Curiously, the growth or decrease can be tuned by choosing different lattice actions. In principle a naive tree-level improvement of the lattice action minimises these lattice artefacts. Nonetheless, one can even use this effect to compensate the flow of the continuum actions towards zero-size instantons with contributions from lattice artefacts~\cite{GarciaPerez:1993lic}. Once the size becomes of the order of the lattice spacing the instanton disappears altogether from the lattice and the topological charge changes along the flow. This phenomenon was observed in the early studies made with cooling~\cite{Teper:1985rb,IWASAKI1983159} but it holds for gradient flow as well.

All these phenomena are present in our studies. Nonetheless, the flow produces very nice smooth configurations revealing the presence of structures larger than the smearing radius. 

To distinguish the definition we denote as $q_c(\tau)$, $S_c(\tau)$ the quantities obtained at flow time $\tau$. If not otherwise specified, the Wilson action is used in the flow and the observable is computed with the clover definition on the flowed observables. Further details of the implementation can be found in~\cite{Luscher:2010iy}.

\subsection{Adjoint filtering method (AFM)}
\label{sec:AFM}
Motivated by the previous considerations it would be interesting to find a filtering method which would produce nice smooth pictures without  distorting the size distribution of instantons and eliminating  
instanton/anti-instanton pairs. 
This led one of the authors and collaborators to present a method based on zero modes and  quasi-zero modes of the adjoint Dirac operator~\cite{Gonzalez-Arroyo:2005fzm, GarciaPerez:2011tx}.  The idea will be explained below.

Given a gauge field which is a solution of the classical equations of motion one can construct a solution of the   Dirac equation in the adjoint representation as follows:
\begin{align}
\psi^a(V,x)=\frac{1}{8} F_{\mu\nu}^a(x)\comm{\gamma_\mu}{\gamma_\nu}V\; 
\end{align}
for $V$ an arbitrary 4-spinor. This can be called the supersymmetric zero-mode (SZM), since it corresponds to the gluino partner in the supersymmetric vector representation.
Using the projectors $P_\pm=(\mathbf{I}\pm \gamma_5)/2$ one can obtain the positive and negative chirality counterparts which are solutions of the left and right-handed Weyl equation respectively. Choosing an appropriate basis these bi-spinorial fields can be written as follows:
\begin{eqnarray}
\psi_+^a\ (x)=-\frac{i}{4} F_{\mu \nu}^a(x) \sigma_\mu \bar{\sigma}_\nu v_+=\frac{1}{2}(E_i(x)+B_i(x))\tau_i v_+ \\
\label{psi_plus}
\psi_-^a\ (x)=-\frac{i}{4} F_{\mu \nu}^a(x) \bar{\sigma}_\mu \sigma_\nu v_- =\frac{1}{2}(-E_i(x)+B_i(x))\tau_i v_- \; ,
\label{psi_minus}
\end{eqnarray}
where $\sigma_\mu$ is a quaternionic $2\times 2$ matrix ($\sigma_\mu=(\mathbf{I}, -i \vec{\tau})$) given in terms of the Pauli matrices ($\tau_i$), and $\bar{\sigma}_\mu$ are their adjoints. Since, $v_\pm$ are arbitrary bi-spinors, the solution runs over a two-dimensional space spanned by one mode and its charge conjugate. Choosing $v_\pm$ of unit norm, the density of these modes satisfy
\begin{align}
S_{AFM}(x)&=\frac{1}{g^2} (|\psi_+|^2 +|\psi_-|^2 )
  = \frac{1}{4g^2} F_{\mu\nu}^a
F_{\mu \nu}^a \\
q_{AFM}(x)&=\frac{1}{8\pi^2}(|\psi_+|^2 - |\psi_-|^2) 
  = \frac{1}{32\pi^2 } F_{\mu\nu}^a
\tilde{F}_{\mu \nu}^a \; .
\label{AFM_top}
\end{align}
When ultraviolet fluctuations are added to the classical solution, the filtering property of these solutions becomes apparent. 
The AFM densities ($S_{AFM}(x)$ and $q_{AFM}(x)$) are obtained from the lowest modes of the adjoint Dirac operator using \eqref{AFM_top}. Despite the modification by the noise, these densities remain close to the original zero-mode values while the values computed from the gauge field directly show large deviations. This has been shown analytically in Ref.~\cite{GarciaPerez:2011tx} and becomes also clear from the results presented in Sec.~\ref{noiseconf}.   

In general a gauge field  configuration can give rise to several different Dirac zero-modes. The  index theorem relates the number of zero-modes in  any representation to the total topological charge as follows:
\begin{equation}
n_+(R)-n_-(R)= \frac{T_R}{T_F} Q = \frac{c_R d_R}{c_F d_F} Q 
\end{equation}
where $n_\pm(R)$ is the number of left/right handed zero modes of the Dirac operator in representation $R$, and $c_R$ and $d_R$ are values of the quadratic Casimir and the dimension of the representation in question 
(For the fundamental representation $R=F$).
Thus, for non-zero topological charge there should be zero-modes even if the gauge field is not a solution of the classical equations of motion. 
One might wonder how this affects the method, given that the SZM  is only a zero-mode for a classical solution, but it just implies that in general near zero-modes have to be considered.
Furthermore, even if the gauge field is a classical solution there are $2NQ=n_+(A)-n_-(A)$ zero-modes in the adjoint representation.  How can one distinguish the SZM within this set?
Notice that the adjoint representation is real and zero-modes (and other eigenstates of the Dirac operator) come in CP pairs just as we saw for the SZM. Indeed, the two Weyl spinors of a CP pair can be combined using a quaternionic $2\times 2$ matrix representation. Hence, the SZM is just one pair within these $NQ$ zero-mode pairs. It is expected to select the most relevant information from the zero modes.

There is one particular feature of the SZM which allows one to distinguish it from other zero or quasi-zero modes of the adjoint Dirac operator. Taking for example $v_+^t=(1,0)$
the positive chirality SZM becomes 
\begin{align}
\psi_+(x)=\frac{1}{2}
\begin{pmatrix}
(E_3(x)+B_3(x))\\
(E_1(x)+B_1(x))/2+i(E_2(x)+B_2(x))/2
\end{pmatrix},
\end{align}
Notice that the first component is real at all points of space. This is a very strong condition not shared by other zero-modes. 
A similar result applies for the negative chirality component $\psi_-(x)$. 
This property was the basic ingredient put forward in Ref.~\cite{Gonzalez-Arroyo:2005fzm} to select the SZM from within other 
(quasi)zero modes.

In Ref.~\cite{GarciaPerez:2011tx} a more elegant way to  define the method was proposed. The left/right
SZM was identified with the eigenvector of lowest eigenvalue of the operators
\begin{equation}
\label{eq:continuum_O}
\Oopm= P_\pm  P_0  (\gamma_5 \Dsla)^2 P_0 P_\pm  
\end{equation}
where $P_0$ is a projector onto the subspace of states satisfying the reality condition, while $P_{\pm}$ 
projects onto the positive and negative chirality sectors. 
 
\subsection{Lattice implementation of the adjoint filtering method (AFM)}
On the lattice the analysis of zero-modes and quasi-zero modes depends strongly upon the choice of lattice Dirac operator $D_L$. 
It is clear that the construction relies strongly on chiral symmetry represented on the lattice by the Ginsparg-Wilson relation. This relation allows a lattice definition of an integer valued  topological charge
using the Dirac operator in the fundamental representation~\cite{Hasenfratz:1998ri}:
\begin{align}
    Q=\frac{1}{2}\text{tr}\Big[\gamma_5\big(2 - aD_L\big)\Big]  = n_+-n_-\;.
\end{align}
For smooth configurations this  coincides with the topological charge obtained from the gauge field up to lattice artefacts.
Similarly, one can also construct a corresponding lattice definition of the topological charge density  as follows~\cite{Niedermayer:1998bi}
\begin{align}
    q(x)=\frac{1}{2a^4}\text{tr}_{\text{CD}}\Big[\gamma_5\big(2 - aD_L(x|x)\big)\Big] 
\end{align}
also consistent with gauge field definition up to lattice artefacts.
This is a truly remarkable result.  
However, in practice, a cut on the number of modes that enters in the trace needs to be done. The cut  can be understood as a certain filtering~\cite{Bruckmann:2006wf, Solbrig:2007nr}. In fact, a small number of modes might be enough to signal the presence of topological structures on the configurations. Nonetheless, the filtering not only reduces the ultraviolet noise but also distorts the structures even for smooth configurations. Zero-modes in the fundamental representation do not have densities that match that of the gauge field. This feature only appears for the SZM in the adjoint representation. 

The cleanest definition of the AFM on the lattice is obtained via the massless overlap Dirac operator ($D_{ov}$)~\cite{Neuberger:1997fp,Neuberger:1998my}. 
\begin{align}
D_{ov}=\frac{1}{2a} (1+\gamma_5 \epsilon(\boldsymbol{H}))\; .
\end{align}
$\boldsymbol{H}=\gamma_5 D_w$ is often referred as the kernel of the Overlap operator and $\epsilon(\boldsymbol{H})$ is the sign function defined through the spectral decomposition of $\boldsymbol{H}$. For the kernel we took $\boldsymbol{H}=\gamma_5 D_w$, where $D_w$ is the standard Wilson-Dirac operator (DW) in the adjoint representation,
\begin{align}
    D_w = 1 - \kappa \big[(1-\gamma_\mu)(V_\mu(x))\delta_{x+\mu,y} +  (1+\gamma_\mu)(V^ \dagger{}_\mu(x-\mu))\delta_{x-\mu,y}\big]\,,
\end{align}
with links $V_\mu(x)$ in the adjoint representation. The  Overlap operator satisfies the Ginsparg-Wilson relation 
\begin{align}
\gamma_5 D_{ov}+D_{ov}\gamma_5=2aD_{ov}\gamma_5 D_{ov}.
\end{align}
This relation implies that $D_{ov}$, in contrast to the Wilson-Dirac operator $D_w$, is a normal operator and implements chiral fermions on the lattice, which leads to a cleaner identification of the zero modes. For this reason it has been used in many studies that tried to identify QCD vacuum structure such as~\cite{Hasenfratz:1998ri,Ilgenfritz:2007xu,Bruckmann:2011ve,Ilgenfritz:2013oda, Hollwieser:2010mj, Hollwieser:2012kb} and in particular earlier studies of the AFM~\cite{bruno_applications_2010,GarciaPerez:2011tx}.

We have tested different rational and polynomial approximations for the approximation $\epsilon(x)$ of the sign function. However, the optimizations of numerical methods for the AFM are not the focus or the current work and 
we decided to use the simple rational approximation~\cite{Neuberger:1998my, Edwards:1998yw}
\begin{align}
    \epsilon(\boldsymbol{H})= \lim_{N\to\infty}\frac{\boldsymbol{H}}{n}\sum_{n=1}^{N}\frac{1}{\boldsymbol{H}^2a_n+\sigma_n}\; ,
\end{align}
with coefficients $\sigma_n=\sin^2{[\pi (N-1/2)/2n]}$ and $a_n=\cos^2{[\pi(N-1/2)/2n]}$.\footnote{To apply $(\boldsymbol{H}^2a_n+\sigma_n)^{-1}$ to a vector $v_n$ we used the multi shift conjugate gradient method. We choose for the order of the sign approximation $N=120$.}

The AFM is based on the lowest eigenmodes of the operator
\begin{align}
 O^\pm=P_0P_\pm(\gamma_5 D_{ov})^2P_\pm P_0\; ,
\end{align}
which is the lattice counterpart of \eqref{eq:continuum_O}. The chirality projections $P_\pm$ can be consistently applied in case of the overlap operator and $P_0$ projects the real part of the first component to zero.

The lowest eigenvalues have been computed using the PRIMME library \cite{PRIMME1, wu2017primmesvds}. The method provides the smallest $N_\text{ev}$ eigenvectors ($\ket{\psi_{\lambda_i}}$) and eigenvalues ($\lambda_i$) with a given accuracy $\delta$ defined by\footnote{The eigenvalues are always assumed to be ordered according to their size $\lambda_1\leq\lambda_2\ldots$.} 
\begin{align}
    \frac{\bra{\psi_{\lambda_i}}\Oopm-\lambda_i \ket{\psi_{\lambda_i}}}{\bra{\psi_{\lambda_i}}\ket{\psi_{\lambda_i}}}<\delta\; .
    \label{delta}
\end{align}

We define the SZM on the lattice as the eigenvector of smallest eigenvalue ($\lambda_1$) of $O^\pm$ assuming that it is sufficiently well separated from the rest of the eigenvalue spectrum.
The vectors obtained from the eigenvalue algorithm are always normalized to one, which means a normalization factor has to be added. 

There is not always a clear gap between a single SZM and the rest of the spectrum of $\Oopm$. In general several lowest eigemodes have to be considered in the construction of the topological charge densities. Therefore we define
for a given $N_\text{cut}$, which is ideally chosen according to a gap in the spectrum,
\begin{align}
  q_{AFM}(x)=q^{(+)}_{AFM}(x)-q^{(-)}_{AFM}(x)\; ,\quad  q^{(\pm)}_{AFM}(x)=\alpha_i^{(\pm)}\sum_i^{N_\text{cut}} |\psi_i^{(\pm)}|^2(x),
  \label{mode_sum}
\end{align}
with appropriate normalization factors $\alpha_i^{(\pm)}$ and $q_{SZM}(x)=q_{AFM}(x)$ for $N_\text{cut}=1$.
As explained later, especially for noisy configurations, allowing $N_\text{cut}>1$ turns out to be essential. 
In the analysis of Monte Carlo generated configurations, it turns out that instead of a fixed number of modes $N_\text{cut}$ a cut according to a fixed maximal size for the eigenvalues $\lambda_i<\lambda_{cut}$ is more feasible. 
The normalization factors can be set to the same value for all modes and this overall normalization can be selected according to the total topological charge of the configuration. In case of the SZM density, we have applied this procedure. When several modes are included, we have found that a good choice for the normalization is $\alpha_i^{(\pm)}=\frac{1}{N_c}$, which we have used in our numerical studies. 

It is important to note that the overlap operator has an additional parameter, which is the mass (or Hopping parameter $\kappa$) in the Wilson-Dirac kernel. This parameter is not relevant in the continuum limit as long as it is in a certain range. In investigations of lowest eigenmodes of the overlap operator in the fundamental representation, it has been found that the $\kappa$ dependence leads to an effective smearing radius \cite{Moran:2010rn}. As we will see, in the AFM, $\kappa$ will need to be tuned correctly, such that the correct number of zero modes are included in the overlap operator.

\section{Semiclassical contributions and twisted boundary conditions}
\label{sec:sclassical_conf}
The main interest of our study is to be able to identify semiclassical structures within Monte Carlo generated configurations. These include normal instantons, but also objects having fractional topological charge known as fractional instantons. 
These objects appear naturally when considering a  torus with twisted boundary conditions (TBC)~\cite{thooft, vanBaal:1984ar, GarciaPerez:2000aiw, Anber:2023sjn}. These are topological sectors that emerge when constructing gauge bundles with gauge group SU(N)/Z(N) over manifolds having non-trivial two-cycles. Physically these sectors correspond to quantized Z(N) fluxes  through these two-cycles. On the torus   one has one element of the center per plane $z_{\mu \nu}=e^{2 \pi i n_{\mu \nu}/N}$. It is convenient to  decompose the antisymmetric matrix $n_{\mu \nu}$ into two 3-vectors in the standard way ($n_{0 i}=k_i$ and $n_{i j} = \epsilon_{i j l}m_l$). 't Hooft realized (see also ~\cite{vanBaal:1982ag}) that twist is related to the topological charge as follows:
\begin{align}
Q+ \frac{\Pf(n)}{N}= Q+ \frac{\vec{k}\cdot \vec{m}}{N} \in \ZZ\; .
\label{index_theorem}
\end{align}
Thus, for non-orthogonal twists ($\vec{k}\cdot \vec{m} \ne 0 \bmod N $) the topological charge is fractional. The self-dual (anti-self-dual) configurations in this bundle are  the so-called {\em fractional instantons} (anti-instantons). Since they have minimum action they can be obtained by  gradient  flow. This process has been shown to converge in the continuum~\cite{Sedlacek:1982cd} and has been used on the lattice to obtain these configurations for different gauge groups, twist tensors and  torus sizes~\cite{GarciaPerez:1989gt, GarciaPerez:1992fj, Montero:2000mv}. We apply the same methods here to generate fractional instantons having minimal action $Q=1/N$. 
Notice that, as implied by the index theorem, they are unique up to translations, but their size is determined by the torus size. However, even if some of the torus periods go to infinity the solution has a stable limit.

Fractional instantons have been seen to play a major role in a semiclassical description of the physics of gauge fields in a $T^3 \times \R$ environment~\cite{GarciaPerez:1993ab,Gonzalez-Arroyo:1995ynx}. They actually contribute to the restoration of rotational invariance,  center-symmetry, and to recover the value of the string tension. We have chosen this regime to test the filtering methods on Monte Carlo generated configurations. We emphasize that, contrary to the fundamental representation case, the Dirac operator in the adjoint representation and its modes have no singularity or multivaluedness in the presence of twisted boundary conditions. Thus the AFM can be applied straightforwardly.

It is important to realize that fractional instantons are local structures which satisfy the classical equations of motion. This means that they appear also in configurations having no twist and vanishing topological charge. This will be very well exemplified in the explicit smooth configurations that we have used to test the AFM method, to be described below.   

\subsection{Constructing configurations}
\label{constructing configurations}
In this subsection we explain the basic procedure employed in generating
test configurations for our analysis. They correspond to  different situations organised with a varying degree of complexity. The simplest examples are smooth configurations
corresponding to solutions of the classical equations of motion having non-zero values of the topological charge $Q$. These are interesting
cases in which the number of zero-modes of the Dirac equation is
dictated by the index theorem~\eqref{index_theorem}. 

As we mentioned before, it is relatively easy to obtain the unit fractional instantons having $Q=1/N$ by  cooling/gradient-flow techniques.
Once they  have been constructed, one can use them  as a building block to
construct more complex examples. One of the obvious ways is by replicating the torus
in any direction. On the lattice the new torus size in $\mu$ direction
becomes $L'_\mu=p_\mu L_\mu$, where $L_\mu$ is the previous torus size
and $p_\mu$ an integer. The new configuration is also a solution of
the classical equations of motion with topological charge $Q'=p_0
p_1p_2 p_3/N$  and with a  new twist tensor is $n'_{\mu \nu}= p_\mu p_\nu
n_{\mu \nu}$. Since $n'$ is defined modulo $N$, with appropriate
choices the new configuration hase no twist $n'=0$ and the topological charge $Q'$ is an integer. 
All these lattice configurations are to a very good approximation self-dual and the action
saturates the Bogomolny bound. 

Another manipulation that can be done uses time reversal. This
operation changes instantons into anti-instantons and viceversa. 
The interesting thing is that one can glue the instanton and
anti-instanton together into a new configuration with larger temporal
torus size in the time-direction. It has vanishing topological charge $Q'=0$ and no twist in the temporal
direction $n'_{0 i}=0$, but the action is twice the one of the original 
fractional instanton. At the time where the reflected configuration is connected, the chromoelectric
field changes sign. Hence smoothness requires this field to be small at these points.
Obviously one can combine these steps to get more general configurations. 

Ultraviolet noise can be added to the configurations applying a number ($N_{\mathrm{MC}}$) of standard heat bath Monte Carlo steps at a given gauge coupling $\beta$. The smaller $\beta$ and the larger $N_{\mathrm{MC}}$
 the more noise is introduced in these heated configurations. In our test we choose the parameters such that the large distance content remains unchanged in this procedure.
 Combining several steps of heating and cooling/gradient flow steps one could also distort the symmetric structures.
 Furthermore, by choosing different lattice actions in the gradient flow one use 
lattice artefacts to our benefit in distorting the size distributions.
The final goal of the paper is, of course the application of the method to thermalized Monte Carlo generated configurations.
In Tab.~\ref{tab:SmoothConfigs} we summarize the configurations used in the present paper. 
    \begin{table} [h!]
    \begin{center}
    \begin{small}
    \begin{tabular}{|c|c|c|c|c|c|c|c|}
         \hline
         Sec. &Group & Lattice size & $Q$ & twist & type & $\kappa $& $\Xi_{AFM}$ \\
         \hline
         \ref{config1} & SU(2) & $8^4$ & $-4$ & $\vec{m}=\vec{k}=0$& smooth & 0.17 & 0.982 \\
         \hline
         \ref{config2} & SU(2) & $16\times 8^3$ & $0$ & $\vec{m}=\{1,1,1\}, \vec{k}=0 $& smooth, I/A pair & 0.16 & 0.989 \\
         \hline
         \ref{config3} & SU(2) & $4\times20^3$ & $1$ & $\vec{m}=\{0,0,1\}$, $\vec{k}=0$  & smooth, caloron & 0.16 & 0.995 \\
         \hline
         \ref{config4} & SU(3) & $40^2\times6^2$ & $0$ & $\vec{m}=\{0,0,2\}, \vec{k}=0 $& smooth, two I/A pair & 0.17 & 0.999\\
         \hline
         \hline
         \ref{noiseconf} & SU(3) &$40^2\times6^2$ & $0$ &$\vec{m}=\{0,0,2\}, \vec{k}=0 $& \ref{config4}+noise &  0.19  & 0.897 \\
         \hline
         MC 1 & SU(2) & $32 \times 4^3$ & - & $\vec{m}=\{1,1,1\}, \vec{k}=0$ & Monte Carlo $\beta=2.44$ & 0.24& 0.685 \\
         \hline
         MC 2 & SU(2) & $64\times 8^3$ & - & $\vec{m}=\{1,1,1\}, \vec{k}=0$ & Monte Carlo $\beta=2.60$ &  0.24 & 0.895 \\
         \hline
    \end{tabular}
    \end{small}
    \end{center}
    \caption{Summary of the configurations used to test the method. $\Xi_{AFM}$ is the cross-correlation computed between $q_{AFM}(x)$ and a smooth reference density, $q_{c}(x)$. In case of the smooth configurations, this is directly computed from the gauge field. For \ref{noiseconf} we used $q_{c}(x)$ from the configuration before the noise has been added. For MC 1 and MC 2, the clover definition at a late flow time $q_{c}(x,\tau'=4)$ is used as a reference. Furthermore, in MC 1 and MC 2 a small amount of gradient flow has been applied before the AFM ($\tau=0.5$). $\Xi_{AFM}$ in the table corresponds to the average of the complete ensembles in this case, see Sec.~\ref{sec:mc_configs}.}
    \label{tab:SmoothConfigs} 
\end{table}

\section{Comparison of numerical methods on tests configurations}
\label{smooth_configurations}
In this section we present results of the AFM for different test configurations representing certain scenarios relevant for the final analysis ordered according to their complexity.  These configurations have a known semiclassical background which can be used for cross-check and verification. This provides also some hints on how the results are affected by the parameters of the method and how to optimize them. The GF is used in these examples only as a tool to manipulate the configurations. We set the parameter  for the convergence of the iterative solver \eqref{delta} to $\delta=10^{-9}$.

In order to estimate the quality of the filtering, the difference between the topological charge density obtained from the gauge field $q_c(x)$ and the one obtained using the AFM $q_{AFM}(x)$ has to be quantified. This can be done using the cross-correlation of the densities determined with two different methods $A$ and $B$ \cite{Bruckmann:2006wf, Moran:2010rn},
\begin{align}
\chi_{AB}(r)=\frac{\sum_{x,y} (q_A(x)-\ev{q_A})(q_B(y)-\ev{q_B})\delta(|x-y|-r)}{\sum_{x,y}\delta(|x-y|-r)}\, .
\end{align}
To remove ambiguities related to the normalization, one can take the ratio of the latter with respect to its geometrical mean 
\begin{align}
\Xi_{AB}=\frac{\chi^2_{AB}(0)}{\chi_{AA}(0)\chi_{BB}(0)}\, .
\label{Xi:def}
\end{align}
$A$ refers to some filtering technique and $B$ always refers to the gauge definition $q_c(x)$ in the following, therefore we use the notation $\Xi_{A}$.  
The quantity signals the correlation between the different densities:
if the distributions obtained with both methods are the same then $\Xi_A=1$, while for completely different distributions $\Xi_A=0$.

\subsection{Smooth configurations}
On smooth configurations with a given semiclassical content, the topological charge density can be directly obtained with the field theoretical definition \eqref{eq:SQField}. This provides an unbiased cross-check, which does not rely on other methods or assumptions. The filtering technique should reproduce the semiclassical background chosen in the construction of the configurations. 
\subsubsection{Eight fractional anti-instantons, $Q = -4$}
\label{config1}
\begin{figure}[h!]
     \begin{subfigure}{0.5\textwidth}
    \includegraphics[width=\textwidth]{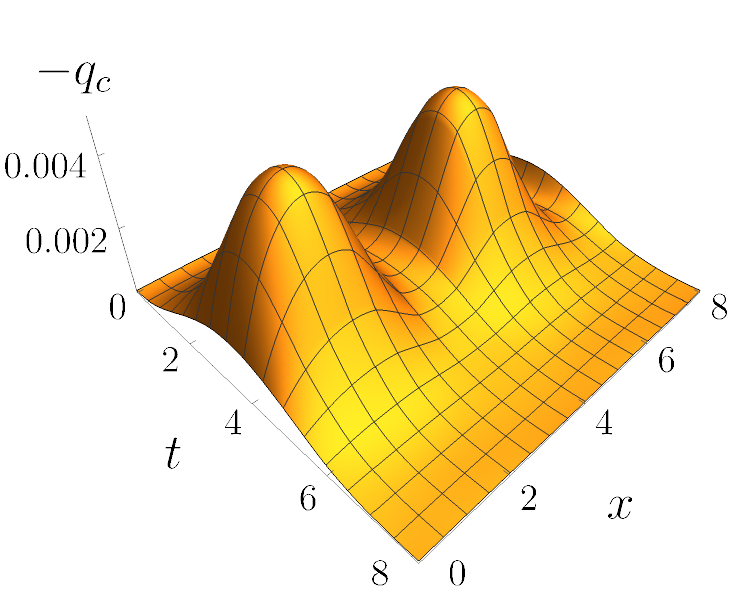}
    \caption{gauge field}
    \label{q4_gauge}
    \end{subfigure}
     \begin{subfigure}{0.5\textwidth}
    \includegraphics[width=\textwidth]{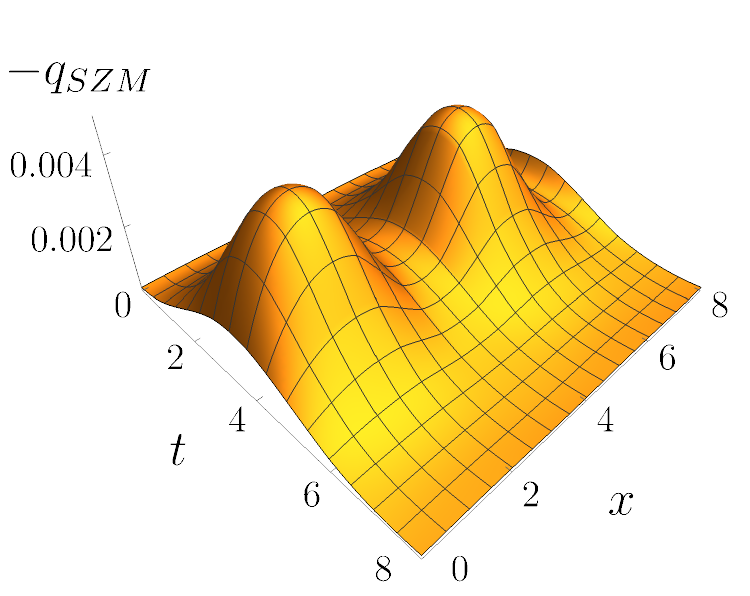}
    \caption{SZM}
    \label{q4_szm}
    \end{subfigure}
    \caption{2D slice of the topological charge density for the $Q=-4$ configuration (Sec.~\ref{config1}) obtained from the gauge field (a) and the supersymmetric zero mode (b). The slice was chosen to cut at the maxima of the two fractional instantons shown. For better visualization, the negative density ($-q$) is shown.}
    \label{fig:Q4}
\end{figure}

 Using the methods explained in Sec.~\ref{constructing configurations}, we have constructed a configuration on a $V=8^4$ lattice with $8$ fractional anti-instantons. We started with an $V=8\times 4^3$ SU(2) fractional anti-instanton and replicated twice every spatial direction eliminating the twist and producing a configuration with $Q_I=-4$ which is to a very good approximation anti-selfdual.  We then  measured the smallest eigenvalues of the chirally projected overlap operator $P_\pm D_{ov} P_\pm$, Tab.~\ref{tab:Overlap_Q4}, and found 16 zero modes in accordance  with the index theorem. This configurations is challenging given the small lattice size occupied by each fractional anti-instanton implying  large lattice artefacts. As a result, the topological charge defined on the lattice by clover plaquettes, $Q_c=-3.5$, deviates significantly from the expected integer value. Nevertheless the topological density defined in this way, still resembles the expected eight fractional anti-instantons. For better visualization, we show 
 in Fig.~\ref{q4_gauge} a two-dimensional slice  of the lattice which captures two of them.

 \begin{table} [h!]
    \centering
    \begin{tabular}{|c|c|c|c|}
        \hline
        {$P_-D_{ov}P_-$} & {$P_+D_{ov}P_+$} & $\Oom$ & $\Oop$\\
        \hline
        $\lambda_{1..16}< 1\cdot 10^{-9} $ & $\lambda_{1,2}= 1.169 \cdot 10^{-1}$ &$\lambda_1=4.993 \cdot 10^{-7}$ & $\lambda_1=1.171 \cdot 10^{-1}$\\
        \hline
        $\lambda_{17,18}= 1.169 \cdot 10^{-1}$ & $\lambda_{3,4} = 1.170 \cdot 10^{-1}$  & $\lambda_2=1.112 \cdot 10^{-3}$ & $\lambda_2=1.171 \cdot 10^{-1}$\\
        \hline
        $\lambda_{19,20} = 1.170 \cdot 10^{-1}$ & $\lambda_{5,6}= 1.430 \cdot 10^{-1}$ & $\lambda_3=1.144\cdot 10^{-3}$ & $\lambda_3=1.172 \cdot 10^{-1}$\\ 
        \hline
        $\lambda_{20,21}= 1.430 \cdot 10^{-1}$ & $\lambda_{7,8}= 2.590 \cdot 10^{-1}$  & $\lambda_4=1.144\cdot 10^{-3}$ & $\lambda_4=1.172 \cdot 10^{-1}$\\
        \hline
    \end{tabular}
    
    \caption{Lowest eigenvalues of the operators $P_\pm D_{ov}P_\pm$ and $\Oopm$ for the configuration presented in Sec.~\ref{config1}. The positive chirality sector, $P_+D_{ov}P_+$ contains no zero modes. Note that the eigenvalues of $P_\pm D_{ov}P_\pm$ are always doubly degenerate.}
    \label{tab:Overlap_Q4}
    \end{table}

In the AFM, we determined the four smallest eigenvalues of the $\Oom$ operator Tab.~\ref{tab:Overlap_Q4}. In this case there is a considerable gap between the smallest eigenvalue, corresponding to the SZM, and the rest of the spectrum. The topological charge density obtained with the AFM reproduces accurately the one of the gauge field definition as shown in Fig.~\ref{q4_szm}. To show the consistency at a more quantitative level, we have performed a parabolic fit around the maxima of each structure according to
\begin{align}
\label{eq:q_fit}
    q_{fit}(x)=h_0-\frac{(\vec{x}-\vec{x}_{0})^2}{\rho_0^2}\; .
\end{align}
The results are summarized in Tab.~\ref{tab:Q4 parameters}.
The fit parameters obtained in this way are indeed very similar which confirms the consistency of the method. Furthermore, the cross-correlation between the two determination is close to one ($\Xi_{AFM}=0.982$).
We also applied the GF for a long flow time $\tau=100$ to confirm the stability of the structures and their character as local minima of the action.

\begin{table}[h!]
    \centering
    \begin{tabular}{|c|c|c|c|c|c|c|}
        \hline
         & x$_0$/a & y$_0$/a & z$_0$/a & t$_0$/a & $\rho_0$/a & h$_0$/a\\
         \hline
         $q_c$ & 1.50 & 2.78 & 0.887& 3.50 & 59.47 & 0.0038\\
         \hline
         $q_{AFM}$ & 1.50 & 2.78 & 0.88 & 3.51& 51.38 & 0.0035\\
         \hline
         $q_c(\tau=100)$ & 1.50  & 2.78 & 0.78  & 3.51 & 50.95 & 0.0038\\
         \hline
    \end{tabular}
    \caption{Parameters obtained by fitting to a parabola (Eq.~\eqref{eq:q_fit}) using 5 points around a particular local maximum for the configuration discussed in Sec.~\ref{config1}. Comparison of the results for the AFM (density of the SZM) and gauge field definition. The result for the large flow time shows the stability of the solutions.}
    \label{tab:Q4 parameters}
\end{table}

\subsubsection{Configuration with a fractional instanton/anti-instanton pair, $Q=0$}
\label{config2}
The starting point for the construction of this configuration was a $Q=1/2$ fractional instanton in a $V=8^4$ lattice with non-orthogonal twisted boundary conditions. We glued this configuration to its time-reversed to produce  a smooth  configuration on a $V=16\times8^3$ lattice which contains a fractional  instanton/anti-instanton pair. This is an interesting configuration to study, since the total topological charge vanishes and it is no longer self-dual or anti-self-dual. Furthermore, it is a situation that must be present in possible realistic descriptions of the Yang-Mills vacuum as 
predicted in the  fractional instanton liquid model \cite{Gonzalez-Arroyo:2023kqv}. 

\begin{figure}[h!]
    \centering
    \includegraphics[width=0.45\linewidth]{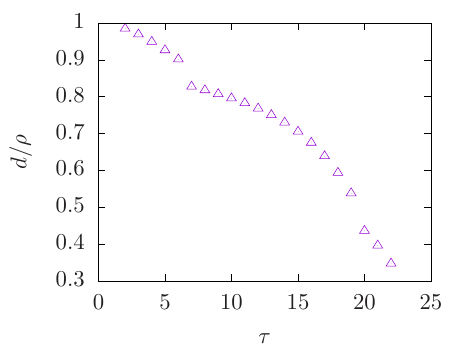}
    \hspace{0.5cm}
    \includegraphics[width=0.45\linewidth]{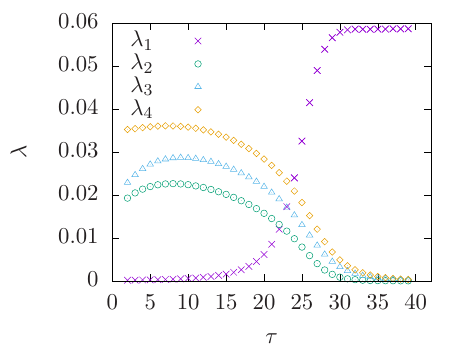}
    \caption{Detailed analysis of the $Q=0$ configuration (Sec.~\ref{config2}). Left: evolution $d/\rho$ during gradient flow. Right: lowest modes of the $\Oopm$ operator as a function of the gradient flow time.}
    \label{fig:Q0_AFM_GF}
\end{figure}

For any smoothing method this kind of configurations presents a major challenge, since the smoothening process tends to erase these structures. 
They do not correspond to minima of the classical action and smoothening methods relying on a minimization of the action, like GF or cooling, lead to important modifications and eventually to the annihilaton of the pair. As mentioned earlier, this is a process that occurs in the continuum and not due to the lattice discretization. Obviously, the rate of annihilation of a pair depends on the ratio of the separation with respect to the size of the objects. This process is illustrated in Fig.~\ref{fig:Q0_AFM_GF}. showing the evolution of the ratio as a function of the GF time. At large  flow times the total action and topological charge density drop rather steeply to very small values approaching the classical vacuum.  It is important to stress that the annihilation is not a lattice artifact but rather a dynamical effect triggered by the GF itself. 

\begin{figure}[h!]
    \centering
    \includegraphics[width=0.45\linewidth]{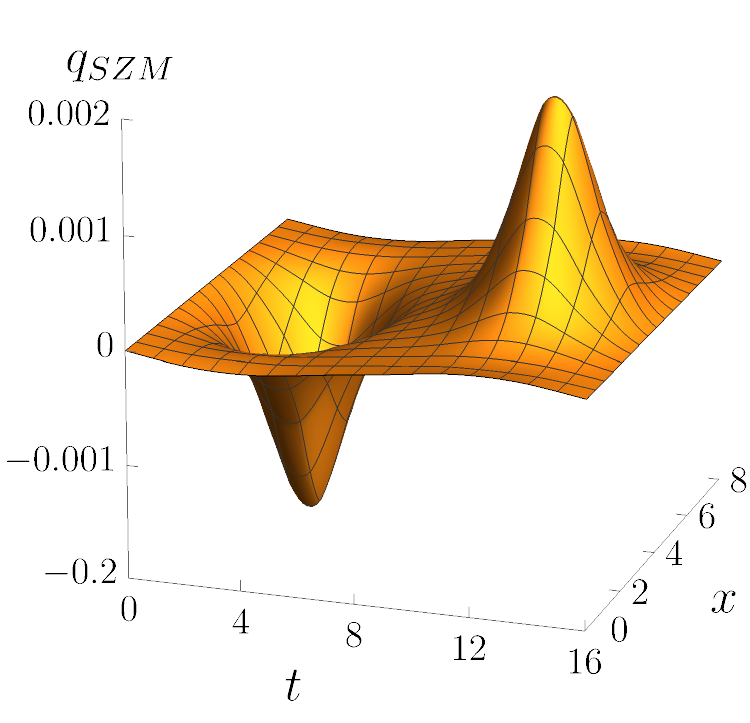}
        \caption{2D slice of the topological charge density obtained with AFM for an instanton/anti-instanton configuration with $Q=0$ (Sec.~\ref{config2}). The slice was chosen to show the maxima of the fractional instantons. Only the lowest mode (SZM) in each chirality is considered.}
    \label{Q0_AFM}
\end{figure}

Also for methods based on the eigenvalues, some additional difficulties are expected due to the absence of zero modes according to the index theorem. However, our results show that the AFM provides a more reliable representation of the structures than the GF. Instead of zero modes, which are excluded by the index theorem, the close to zero modes are reproducing the density of each one of the fractional (anti-)instantons, Fig.~\ref{Q0_AFM}. We computed the spectrum of the projection operator $\Oopm$ and the chirally projected overlap operator $P_\pm D_{ov} P_\pm$ for each chirality sector and found a gap of $\mathcal{O}(10^{-2})$ between the smallest eigenvalues and the rest of the spectrum for both operators Tab.~\ref{tab:AFM Q0}. This is considerably smaller than in the example presented in Sec.~\ref{config1}, but nevertheless sufficient to obtain a reliable result with $\Xi_{AFM}=0.989$. 

One can apply the AFM method to the full sequence of configurations generated by GF.   This allows us to study  how well the AFM is able to discriminate the two structures along the annihilation process. For that purpose we have computed the lowest eigenvalues of $\Oopm$ up to a flow time $\tau=40$, see Fig.~\ref{fig:Q0_AFM_GF}. The SZM captures the instanton/anti-instanton pair until $\tau\simeq22$ where there is a separation of around $d/\rho=0.4$ between them. 
At that point, the gap in the eigenvalue spectrum of $\Oopm$  has dropped to zero and there is a crossing of levels. This is clear since now the lowest state does show a completely different density, but the first excited state displays a certain continuity in shape with the instanton structure.

\begin{table} [h!]
\centering
    \begin{tabular}{|c|c|c|c|}
        \hline
        $\Oom$ & $\Oop$ & $P_-D_{ov}P_-$ & $P_+D_{ov}P_+$ \\
        \hline
        $\lambda_1= 1.17 \cdot 10^{-4}$&$\lambda_1= 1.17 \cdot 10^{-4}$ &$\lambda_{1,2} = 1.03 \cdot 10^{-4}$ &$\lambda_{1,2}=  1.03 \cdot 10^{-4}$ \\
        \hline
        $\lambda_2= 1.75 \cdot 10^{-2}$&$\lambda_2= 1.75 \cdot 10^{-2}$&$\lambda_{3,4} = 1.51 \cdot 10^{-2}$ &$\lambda_{3,4}= 1.51 \cdot 10^{-2}$ \\
        \hline
        $\lambda_3= 2.05\cdot 10^{-2}$&$\lambda_3=  2.05\cdot 10^{-2}$ &$\lambda_{5,6} = 2.47 \cdot 10^{-2}$&$\lambda_{5,6}=2.47 \cdot 10^{-2}$\\
        \hline
        $\lambda_4= 2.77\cdot 10^{-2} $&$\lambda_4= 2.77\cdot 10^{-2}$&$\lambda_{7,8} = 6.83 \cdot 10^{-2}$&$\lambda_{7,8}=6.83 \cdot 10^{-2}$ \\
        \hline
    \end{tabular}
    \caption{Lowest eigenvalues of the $\Oopm$ and $P_\pm D_{ov}P_\pm$ operators for the configuration discussed in Sec.~\ref{config2}.}
\label{tab:AFM Q0}
\end{table}

\subsubsection{A caloron configuration}
\label{config3}
Calorons are solutions of the classical equations of motion for $R^3\times S^1$. This is an interesting case for which there are analytic formulas for the $Q=1$ case which display the decomposition into separate fractional topological charge objects~\cite{Kraan:1998kp,Kraan:1998sn,Lee:1998bb,Lee:1998vu}. As shown in Ref.~\cite{GarciaPerez:1999hs} these can be approximated on the lattice by configurations with one direction much shorter than the others. In this particular case, we used twisted boundary conditions to produce a fractional instanton on a lattice of size $V=20 \times 20 \times 10\times 4$ and replicated the z-direction to produce a  $Q=1$ configuration on a $V=20^3\times 4$ lattice. 

\begin{figure}[h!]
    \centering
    \includegraphics[width=0.45\linewidth]{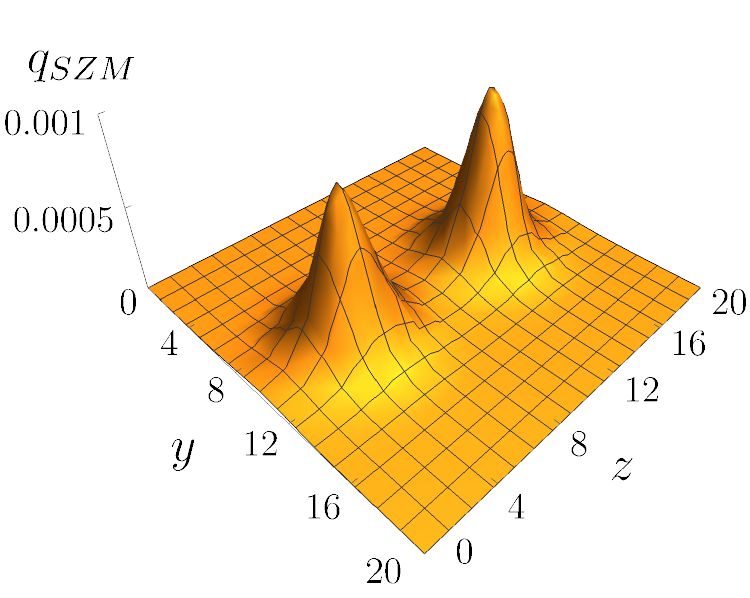}
        \caption{2D slice of the topological charge density of the AFM applied to a $Q=1$ caloron configuration (Sec.~\ref{config3}). Only the lowest mode with positive chirality (SZM) is considered.}
    \label{Caloron}
\end{figure}

The lowest part of the spectrum of the $P_\pm D_{ov}P_\pm$ operator revealed four zero modes with positive chirality Tab.~\ref{tab:Overlap_cal}, which is consistent with the index theorem (\ref{index_theorem}). To apply the AFM, we computed the four lowest modes of the $\Oopm$ Tab.~\ref{tab:Overlap_cal}. We found a gap of order $\mathcal{O}(10^{-3})$ on the $\Oop$ operator between the smallest eigenvalue corresponding to the SZM and the rest of the spectrum. The AFM topological density obtained through (\ref{AFM_top}) reproduced successfully the two separate fractional topological charge objects, as we can see from Fig.~\ref{Caloron} and from the value of the cross-correlation to the clover definition with the topological charge density $\Xi_{AFM}=0.995$.
 \begin{table} [h!]
    \centering
    \begin{tabular}{|c|c|c|c|}
        \hline
        $\Oom$ & $\Oop$ & $P_-D_{ov}P_-$ & $P_+D_{ov}P_+$ \\
        \hline
        $\lambda_1=3.22\cdot 10^{-3} $ &  $\lambda_1= 5.84\cdot 10^{-7}$ &$\lambda_{1,2}= 3.21 \cdot 10^{-3}$ &$\lambda_{1,2}< 1\cdot 10^{-9}$ \\
        \hline
        $\lambda_1=3.22\cdot 10^{-3} $ & $\lambda_2= 7.75\cdot 10^{-4}$ &$\lambda_{3,4}=6.32\cdot 10^{-2}$ &$\lambda_{3,4}<1\cdot 10^{-9}$\\
        \hline
        $\lambda_1=3.22\cdot 10^{-3} $ & $\lambda_2= 7.96\cdot 10^{-4}$ &$\lambda_{5,6}=6.34 \cdot 10^{-2}$  &$\lambda_{5,6}=3.21\cdot 10^{-3}$\\
        \hline
         $\lambda_1=6.32\cdot 10^{-3} $  & $\lambda_2= 7.99\cdot 10^{-4}$  & $\lambda_{7,8}=6.37 \cdot 10^{-2}$  & $\lambda_{7,8}=6.32 \cdot 10^{-2}$ \\
        \hline
    \end{tabular}
    \caption{Lowest eigenvalues of the $\Oopm$ and $P_\pm D_{ov}P_\pm$ operators for the configuration discussed in Sec.~\ref{config3}.}
    \label{tab:Overlap_cal}
    \end{table}
\\

\subsubsection{SU(3) configuration with several fractional instanton/anti-instantons $Q=0$}
\label{config4}
\begin{figure}[h!]
    \centering
    \includegraphics[width=0.45\linewidth]{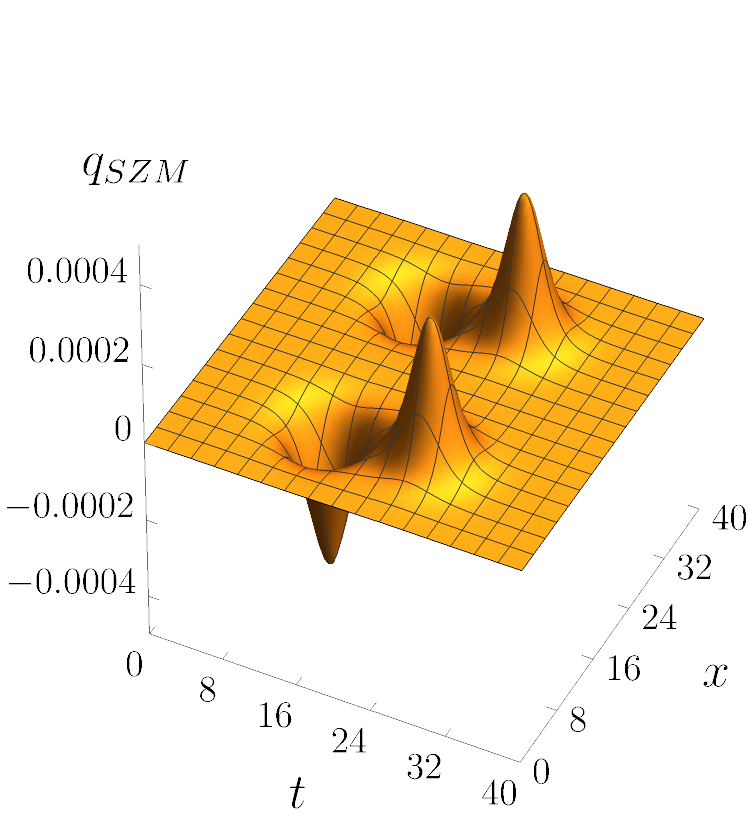}
    \hspace{1cm}
        \caption{Topological charge density (2D slice) obtained with AFM for a two instanton/anti-instanton pair configuration based on SU(3) gauge group with $Q=0$ (Sec.~\ref{config2}). Only the lowest mode (SZM) in each chirality is considered.}
    \label{fig:40x6_smooth}
\end{figure}

In order to test the method also for larger gauge groups, we present as the last smooth example an SU(3) configuration on a $V=6\times6\times40\times40$ lattice with $Q=0$. This type of lattice geometry approximates the $T^2\times R^2$ case. The configuration contains two fractional instantons and two fractional anti-instantons. This is obtained by gluing along the temporal direction one configuration containing two fractional instantons and it's time-reversed image containing two anti-instantons. 
Like discussed in Sec.~\ref{config2}, this presents a non-trivial case due to possible annihilation of the instantons. The quasi-zero modes corresponding to the fractional instanton structures are clearly distinguishable for the overlap spectrum Tab.~\ref{tab:40x6_smooth_spectrum} and there is a gap of order $\mathcal{O}(10^{-3})$ of the lowest eigenvalues of the $\Oopm$ operator. 
Also in this case the SZM density managed to reproduce the topological content present of the configuration Fig.~(\ref{fig:40x6_smooth}), presenting a high cross-correlation $\Xi_{AFM}=0.999$.
\begin{table} [h!]
    \centering
    \begin{tabular}{|c|c|c|c|}
        \hline
        $\Oom$ & $\Oop$ & $P_-D_{ov}P_-$ & $P_+D_{ov}P_+$ \\
        \hline
        $\lambda_1= 4.79 \cdot 10^{-5}$ & $\lambda_1= 4.79 \cdot 10^{-5}$ &$\lambda_{1,2} = 4.79 \cdot 10^{-5}$ &$\lambda_{1,2}=  4.79 \cdot 10^{-5}$ \\
        \hline
        $\lambda_2= 4.95 \cdot 10^{-5}$ & $\lambda_2= 4.95 \cdot 10^{-5}$&$\lambda_{3,4} = 4.94 \cdot 10^{-5}$ &$\lambda_{3,4}= 4.94 \cdot 10^{-5}$ \\
        \hline
        $\lambda_3= 2.16\cdot 10^{-2}$&$\lambda_3=  2.16\cdot 10^{-2}$ &$\lambda_{5,6} = 2.98 \cdot 10^{-2}$&$\lambda_{5,6}=2.98 \cdot 10^{-2}$\\
        \hline
        $\lambda_4= 2.16\cdot 10^{-2} $&$\lambda_4= 2.16\cdot 10^{-2}$&$\lambda_{7,8} = 3.1 \cdot 10^{-2}$&$\lambda_{7,8}=3.1 \cdot 10^{-2}$ \\
        \hline
    \end{tabular}
    \caption{Lowest eigenvalues of the $\Oopm$ and $P_\pm D_{ov}P_\pm$ operators for the configuration discussed in (Sec.~\ref{config3}).}
    \label{tab:40x6_smooth_spectrum}
\end{table}

\subsection{Filtering of semiclassical background from noisy configurations}
\label{noiseconf}
The first check on smooth configurations has confirmed that the method reproduces well the semiclassical features. As a next step, we test the filtering properties of the AFM method. The aim is to distinguish the semiclassical background from ultraviolet noise. 

\begin{figure}[h!]
    \begin{subfigure}{0.45\textwidth}
    \includegraphics[width=\textwidth]{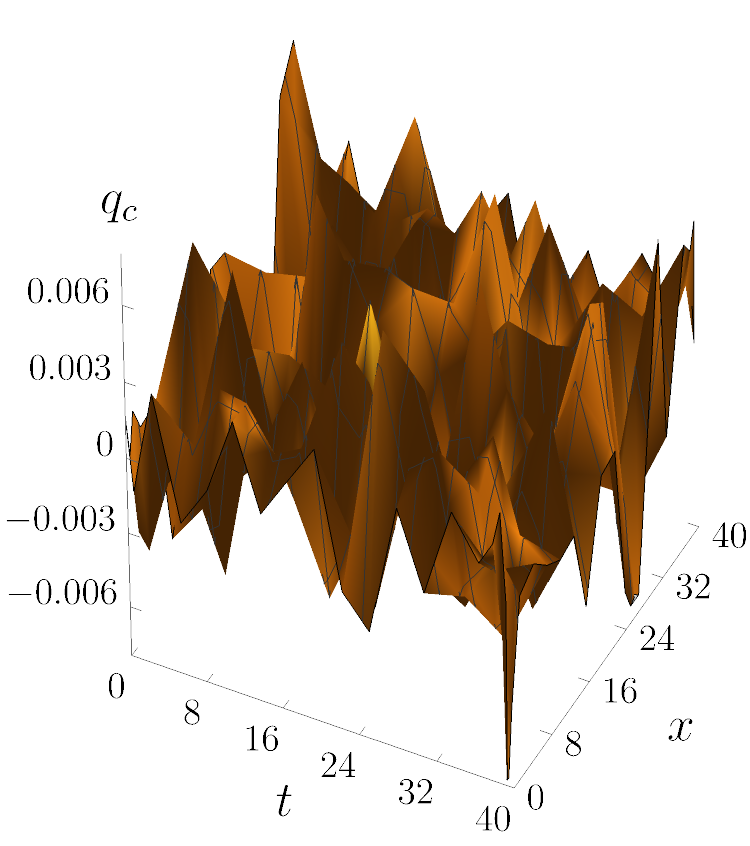}
    \caption{gauge field}
    \end{subfigure}
    \hspace{0.5cm}
    \begin{subfigure}{0.45\textwidth}
        \includegraphics[width=\textwidth]{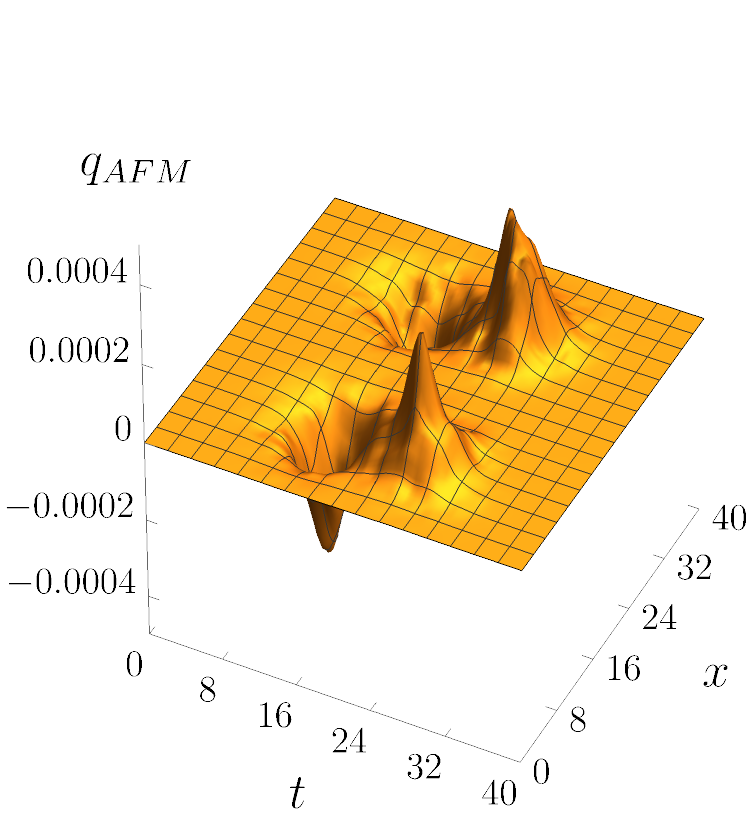}
        \caption{AFM}
    \end{subfigure}
    \caption{Topological charge density (2D slice) of the heated configuration discussed in Sec.~\ref{noiseconf}. (a) gauge field definition, (b) density obtained with the AFM summing two modes in each sector.}
    \label{fig:40x6_noise}
\end{figure}

We take the smooth configuration analyzed in Sec.~\ref{config4} and add noise to it. This is done by applying 20 standard heat bath sweeps based on the plain Wilson action at $\beta=3.2$.
Afterwards, the noise has erased all signs of the semiclassical background  if the topological density is computed directly from the gauge field of the ``heated'' configuration, see  Fig.~\ref{fig:40x6_noise}.

\begin{figure}
    \centering
    \includegraphics[width=0.40\linewidth]{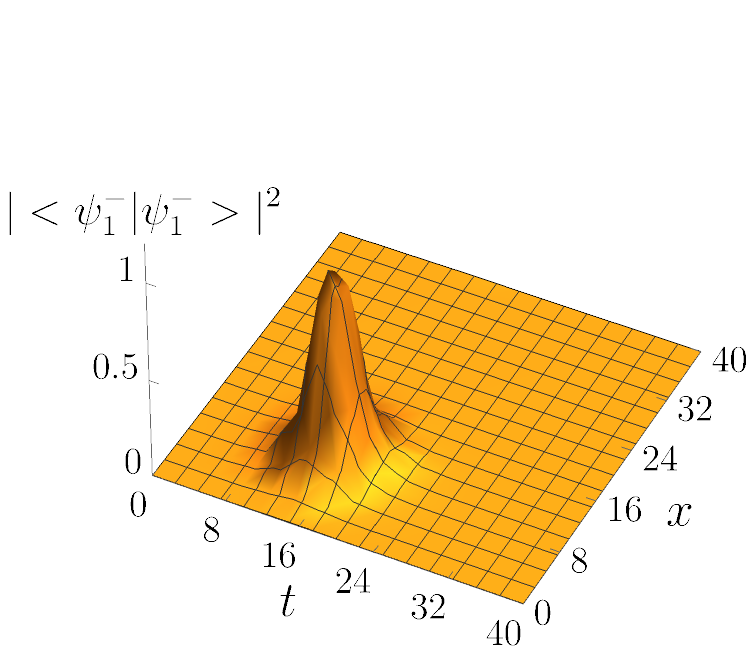}
    \hspace{1cm}
    \includegraphics[width=0.40\linewidth]{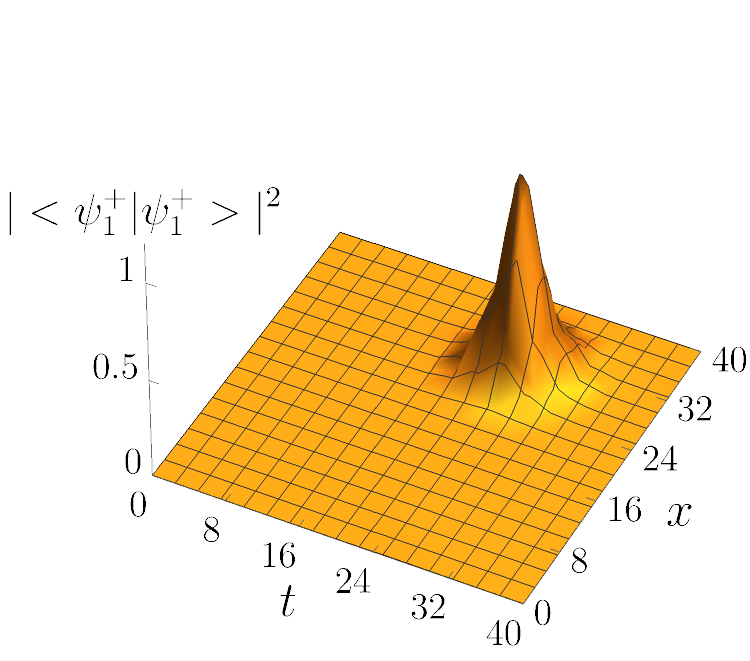}
    \includegraphics[width=0.40\linewidth]{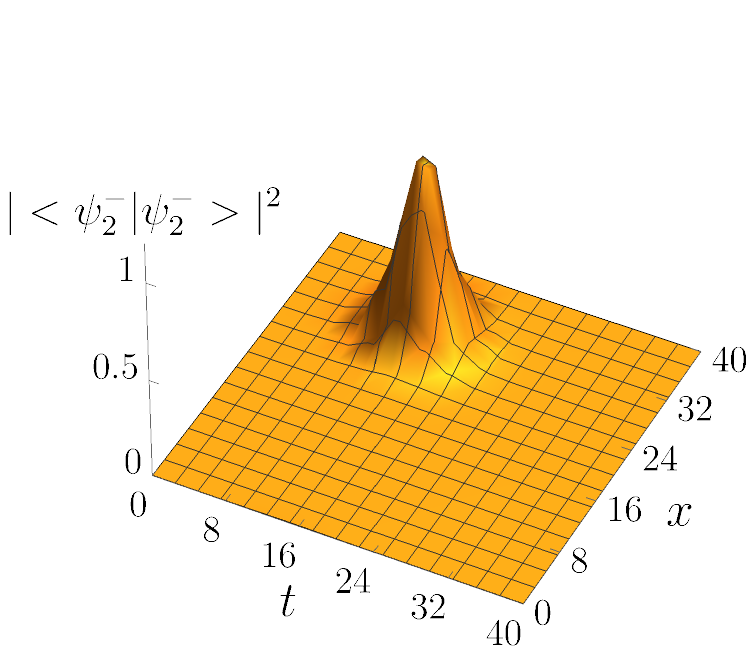}
    \hspace{1cm}
    \includegraphics[width=0.40\linewidth]{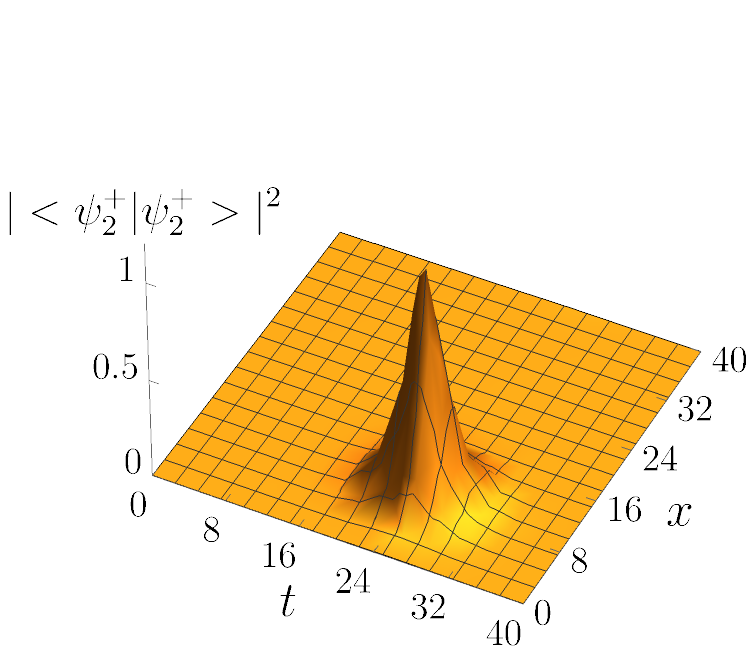}
    \caption{Density obtained from the two lowest modes of operators $\Oom$ (left) and  $\Oop$ (right).}
    \label{fig:40x6_heated_afm}
\end{figure}

In comparison to the smooth configuration, the gap separating the two smallest eigenvalues $\Oopm$ in each sector and the rest of the spectrum is drastically reduced, see Tab.~\ref{tab:40x6_heated_spectrum}. These smallest two eigenvalues remain nearly degenerate. Plotting the densities generated from each of the corresponding eigenvectors separately, as shown in Fig.~\ref{fig:40x6_heated_afm}, revealed that each mode peaks in only one of the four factional structures. 
This is different from the smooth configuration before the noise has been added, Sec.~\ref{config4}. The property of the SZM that it provides an equally weighted average of the contributions from each structure can hence be lost once the noise is added. We have made similar observations also for the configurations generated from Monte Carlo simulations. In several cases each individual lowest eigenmode contributing to $q_{AFM}(x)$ in Eq.~\eqref{mode_sum} distinguishes between the different fractional topological objects which build up the complete topological charge density. This can, of course, only be valid in an approximate limit since, due to the non-linearity, $q(x)$ can not be decomposed into a simple sum of fractional contributions.
In Fig.~\ref{fig:40x6_noise} we show, how the complete AFM produces a clear filtered image re-obtaining the topological density of the smooth configuration and, consequently, the cross-correlation to the topological charge of the original smooth configuration before heating is high $\Xi_{AFM}=0.897$. This shows that the AFM is indeed a filtering method that can eliminate the UV fluctuations introduced by Monte Carlo simulation once the contribution from a sufficient number of modes is combined.

\begin{table}  
    \centering
    \begin{tabular}{|c|c|c|c|}
        \hline
        $\Oom$ & $\Oop$ & $P_-D_{ov}P_-$ & $P_+D_{ov}P_+$\\
        \hline
        $\lambda_1= 3.32 \cdot 10^{-2}$ & $\lambda_1= 3.38 \cdot 10^{-2}$ & $\lambda_{1,2}= 2.60 \cdot 10^{-4}$ & $\lambda_{1,2}= 2.60 \cdot 10^{-4}$   \\
        \hline
        $\lambda_2= 3.6\cdot 10^{-2}$ & $\lambda_2= 3.56 \cdot 10^{-2}$& $\lambda_{2,3}= 5.18 \cdot 10^{-4}$ & $\lambda_{2,3}= 5.18 \cdot 10^{-4}$   \\
        \hline
        $\lambda_3= 8.5\cdot 10^{-2}$&$\lambda_3=  9.26\cdot 10^{-2}$ & $\lambda_{3,4}= 9.86 \cdot 10^{-2}$ & $\lambda_{3,4}= 9.86 \cdot 10^{-2}$   \\
        \hline
        $\lambda_4= 8.7\cdot 10^{-2} $&$\lambda_4= 9.5\cdot 10^{-2}$ & $\lambda_{5,6}= 1.04 \cdot 10^{-1}$ & $\lambda_{5,6}= 1.04 \cdot 10^{-1}$   \\
        \hline
    \end{tabular}
    \caption{Lowest eigenvalues of the $\Oopm$ and the $P_{\pm}D_{ov}P_\pm$ operator for the heated configuration discussed in Sec.~\ref{noiseconf}.}
    \label{tab:40x6_heated_spectrum}
\end{table}

Before we discus the results on the Monte Carlo configurations we collect the main results we obtained so far on smooth configurations
\begin{itemize}
    \item The AFM reproduces correctly the topological density of self-dual smooth configurations. In particular, only the lowest mode of the $\Oopm$, the SZM, is needed.
    \item For smooth configurations which are not self-dual, i.e instanton/anti-instanton pairs, the AFM also reproduces the topological density without altering the configuration.
    \item In the presence of noise, a hierarchy on  the spectrum of the $\Oopm$ appears. The density of the (quasi)zero modes distinguish the different fractional instantons present on the configuration. Adding the densities of all (quasi)zero modes reproduces correctly the topological density.
\end{itemize}

\section{Application in Monte Carlo simulations}
\label{sec:mc_configs}
In a final test, we apply the AFM to ensembles of Monte Carlo generated configurations.
It is expected that these contain a combination of all phenomena discussed so far. The noise level also implies a less obvious dependence on the parameters of the method.
Especially on a coarse lattice, an additional improvement of the Dirac operator might be required. The simplest improvement is a smearing of the link fields in the Dirac operator.
We consider therefore a short GF before applying the AFM. The flow time should in this case be chosen such that the smearing radius remains below a few lattice spacings. As discussed below, this increases the gap in the spectrum of $\Oopm$ and helps to distinguish the doublers from the physical branch in the spectrum of the Wilson-Dirac kernel. In that way, the parameters of the method can be tuned more easily and a much better performance can be achieved.

We study the combination of the AFM and the GF also at larger flow times. Such a study interpolates between a the noisy Monte Carlo generated and an ensemble of smooth configurations. The AFM performs very well on these smooth configurations, even though they contain less separated and symmetric structures than examples discussed in the last sections.
\subsection{Yang-Mills theory in a twisted box}
In these first studies we aim for a setting with a nontrivial semiclassical dynamics on rather small lattices in order to repeat the calculation with different parameters at a reasonable computational cost.
In addition, the setup should allow to control the density and size of typical structures. 
For this reason we considered a SU(2) Yang-Mills theory in a small twisted box like previously studied in Ref.~\cite{GarciaPerez:1993ab}. It corresponds to a $T^3\times R$ geometry with a spatial twist   $m=\{1,1,1\}$, which is fully symmetric in all $T^3$ directions. On the lattice this is implemented by a large ``time'' direction representing $R$ and small ``spatial'' $T^3$ directions. An additional advantage of this setting is that the relevant features of the densities can be shown as a function of the large time direction after  integrating  over the spatial coordinates.

In the following we discuss  some features of this setting, see Ref.~\cite{GarciaPerez:1993ab} for further details. The potential energy has two minima connected by a non-trivial singular gauge transformation. The order parameter distinguishing the two vacua is the Polyakov loop winding once in every spatial direction $P_{(0,1,1,1)}$. The Polyakov line in one given direction $\mu$ is defined by the product of all links in that direction
\begin{align}
\label{Polyakov}
P_\mu(x_\bot)=\prod_{x_\mu=1}^{N_\mu} U_\mu(x_\mu,x_\bot)\; ,
\end{align}
where $x_\bot$ are the coordinates orthogonal to direction $\mu$. Different Polyakov loops are traces of products of $P_\mu$, in particular
\begin{align}
P_{(a,b,c,d)}(x_{\bot})=\frac{1}{N_c}\Tr\left[ P_0^a(x_{\bot}) P_1^b(x_{\bot}) P_2^c (x_{\bot})P_3^d(x_{\bot})\right] \; ,
\end{align}
where $x_{\bot}$ is now orthogonal to all directions with a nontrivial winding of the loop, i.~e.\ $P_{(0,1,1,1)}$ only depends on the time coordinate $x_0$. Indeed, the same features hold if the straight line Polyakov loop (Eq.~\eqref{Polyakov}) is replaced by other paths having the same winding. These are often more useful since they have smaller statistical fluctuations. 

Choosing one of the two vacua corresponds to the breaking of a Z(2) subgroup of center symmetry. The situation resembles the double-well potential in one dimension. It is easy to see that the minimum action configuration that tunnels between the two vacua is precisely a $Q=1/2$ fractional instanton (or anti-instanton). This has a finite free-energy, thus restoring the center symmetry.  This free energy grows with  the physical size of the spatial torus. In a lattice Monte Carlo simulation of the system, the size depends on the  spatial  size of the lattice and on $\beta$. At larger values of $\beta$ the probability to generate fractional instantons decreases and their average number per configuration is small. 
In this regime these topological structures are identified more easily and a semiclassical calculation of the observables is possible. 

For our study we have chosen a regime with moderate values of $\beta$ which provides already a nontrivial challenge for the detection and pairs of instanton-anti-instantons appear. 
We have also chosen in our first tests a rather small lattice to reduce the numerical costs of the method since we want to test larger numbers of eigenmodes and different parameters. The small lattice, however, leads to structures, which are less smooth compared to a setup closer to the continuum limit. More realistic cases would likely provide higher densities but also finer lattice spacings. In order to check the scaling of the results, we have considered two different lattice spacing.

Our concrete parameters are a coarse lattice of $V=4^3\times 32$ and $\beta=2.44$ and a finer lattice of $V= 8^3\times 64$ and $\beta=2.60$. According to our rough scaling estimates they have the same physical lattice sizes but different lattice spacings. For each lattice we have selected a subset of 100 thermalized configurations well separated by a number of 1000 and 2000 heatbath sweeps for the coarse and finer lattice respectively. 
\subsection{Eigenvalue spectrum of Wilson-Dirac operator and $\Oopm$ operator}
\begin{figure}[h!]
\begin{subfigure}{0.5\textwidth}
\includegraphics[width=\textwidth]{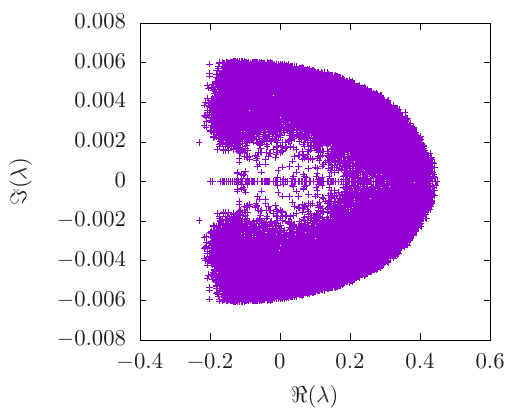}    
\caption{$\tau=0$}
\label{fig:wilson_mc_tau0}
\end{subfigure}
\begin{subfigure}{0.5\textwidth}
\includegraphics[width=\textwidth]{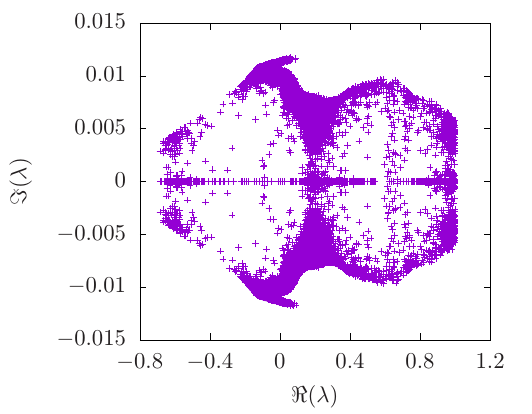}
\caption{$\tau=0.5$}
\end{subfigure}
\caption{Lowest part of the spectrum of the DW operator for 100 Monte Carlo generated configurations with our coarse lattice $V=4^3 \times 32$ ($\beta=2.44$, $\kappa=0.24$) at two different flow times.}
\label{wilson_mc}
\end{figure}

\begin{figure}[h!]
\centering
\includegraphics[width=0.49\linewidth]{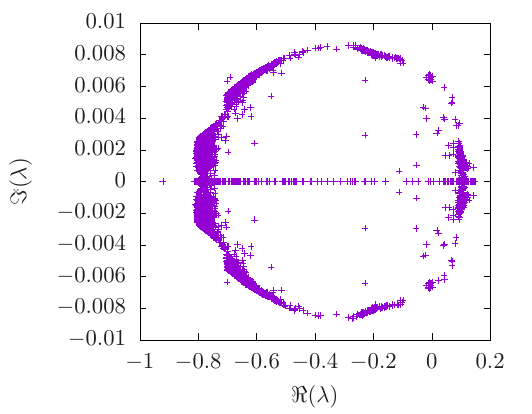}
\caption{Lowest part of the spectrum of the DW operator for 100 Monte Carlo generated configurations, lattice size $V=8^3\times 64$, $\beta=2.60$ at flow time $\tau=0.5$ and $\kappa=0.24$.}
\label{wilson_mc_8x64}
\end{figure}

To tune the $\kappa$ parameter we computed the low-lying part of the spectrum of the Wilson-Dirac operator in the adjoint representation, see Figures~\ref{wilson_mc} and \ref{wilson_mc_8x64}. This already shows a highly nontrivial topological activity with real modes spread along the real axis. It is hence difficult to distinguish within the real modes those that are doublers, but the situation is significantly improved after GF with a small flow time is applied.
Based on these findings we have chosen $\kappa=0.24$ as it seems reasonably aligned with a gap in the spectrum of the Wilson-Dirac operator.

\begin{figure}[h!]
\begin{subfigure}{.5\textwidth}
\includegraphics[width=\textwidth]{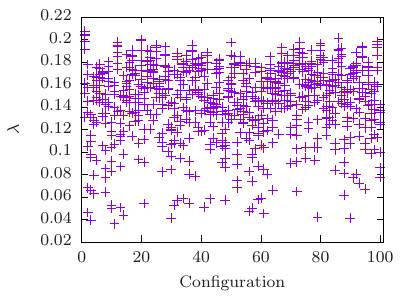}
\caption{$\tau=0$}
\end{subfigure}
\begin{subfigure}{.5\textwidth}
\includegraphics[width=\textwidth]{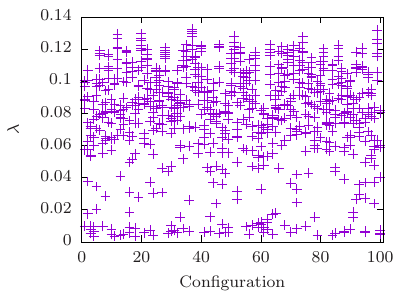}    
\caption{$\tau=0.5$}
\end{subfigure}
\caption{Lowest eigenvalues of $\Oopm$ after GF with flow time $\tau$ for 100 Monte Carlo generated configurations with $V=4^3\times 32$ at $\beta=2.44$.}
\label{O_spectrum}
\end{figure}

\begin{figure}[h!]
\begin{subfigure}{.5\textwidth}
\includegraphics[width=\textwidth]{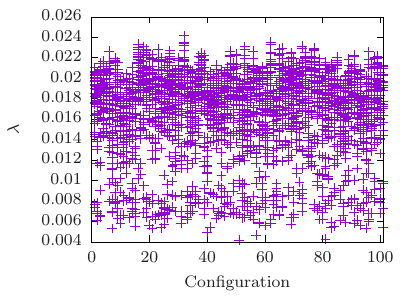}
\caption{$\tau=0.5$}
\end{subfigure}
\begin{subfigure}{.5\textwidth}
\includegraphics[width=\textwidth]{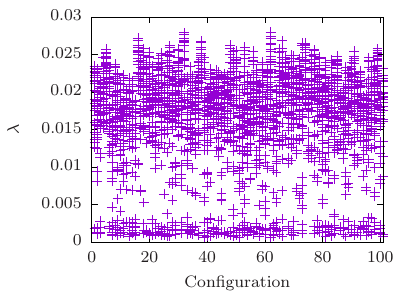}
\caption{$\tau=2.0$}
\end{subfigure}
\caption{
Lowest eigenvalues of $\Oopm$ after GF with flow time $\tau$ for 100 Monte Carlo generated configurations with $V=8^3\times 64$ at $\beta=2.60$.}
\label{O_spectrum_8x64}
\end{figure}

The effect of the nonzero flow time on the spectrum of the operator $\Oopm$ can be seen in Figures~\ref{O_spectrum} and \ref{O_spectrum_8x64}. A cluster of low eigenvalues separated from a more dense region at higher values can be identified. The gap separating these two increases after some GF is applied. It seems that a suitable cut in the spectrum $\lambda_{cut}$ can be determined in order to single out the states that should enter for the computation $q_{AFM}(x)$ in Eq.~(\ref{mode_sum}). This is quantified in more detail in the following sections.

\subsection{Results on the coarse lattice}
At first we discuss in detail the ensembles on the coarse lattice with $V= 4^3\times 32$ at $\beta=2.44$.
In the following we need to quantify the accuracy in order to optimize the parameters. The best way for a quantitative comparison that we have found is in terms of the correlation \eqref{Xi:def} with $q_c(x,\tau')$, obtained at a certain reference flow time $\tau'$. We have chosen $\tau'=4$ since the UV noise is sufficiently filtered out at that flow time and the configurations are rather smooth. At the end we will discuss some possible shortcomings of this approach.

\begin{figure}[h!]
\centering
\includegraphics[width=0.65\linewidth]{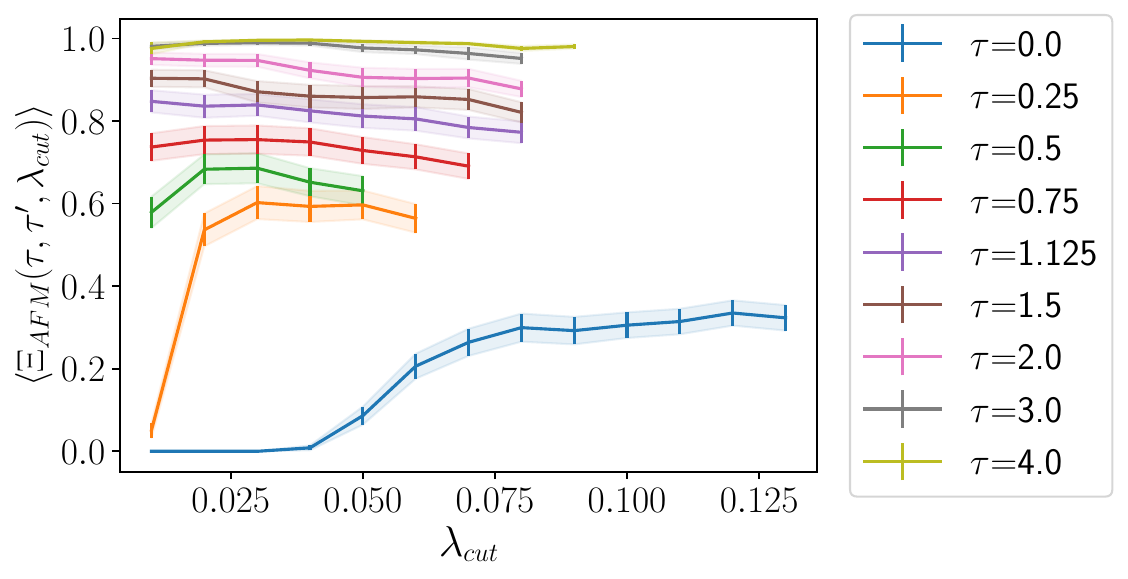}
\caption{Average correlation $\langle\Xi_{AFM}(\tau,\tau',\lambda_{cut})\rangle$ between AFM density at $\tau$ and $q_c$ obtained at a large flow time $\tau'=4$. The values are shown as a function of the cut on the sum of the eigenvalues \eqref{mode_sum}. The lines end where for one of the configurations $\lambda_{cut}$ is larger than the largest computed eigenvalue. If there is no eigenvalue below $\lambda_{cut}$ for a configuration,  $\Xi_{AFM}(\tau,\tau',\lambda_{cut})$ is set to zero.}
\label{Xi}
\end{figure}

An important parameter of the AFM is the cut $\lambda_{cut}$ on the number of eigenmodes used in \eqref{mode_sum}. A second relevant parameter is the flow time for the initial GF applied before the AFM. Hence we define $q_{AFM}(x,\tau)$ as the density obtained with the AFM after GF with flow time $\tau$ has been applied. Then, $\langle\Xi_{AFM}(\tau,\tau',\lambda_{cut})\rangle$ represents the average correlation of this density with $q_c(x,\tau')$ at $\tau'\ge\tau$. 
The dependence on these two parameters is shown in Fig.~\ref{Xi}, where each line of points represents the dependence on $\lambda_{cut}$ at a fixed $\tau$. In the smooth case at $\tau=4.0$ the information is contained in the near zero modes agrees well with the field theoretical definition. At lower $\tau$ the optimal $\lambda_{cut}$ remains small. In case of $\tau=0$ and $\tau=0.25$, the relevant information seems to be spread over a larger eigenvalue region. This observation can also be understood from Fig.~\ref{fig:wilson_mc_tau0}: without GF, the lowest eigenvalues of $\Oopm$ are significantly larger and there is no clear separation from the rest of the spectrum.

\begin{figure}[h!]
\centering
\includegraphics[width=0.50\linewidth]{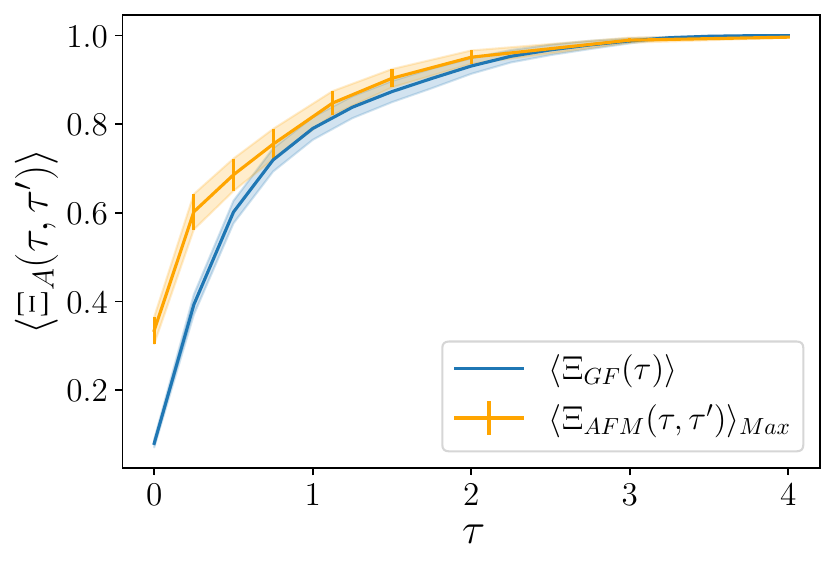}
\caption{Maximum $\langle\Xi_{AFM}(\tau,\tau')\rangle_{max}$ with respect to the cut $\lambda_{cut}$ compared to $\langle\Xi_{GF}(\tau,\tau')\rangle$.}
\label{GF_AFM}
\end{figure}

For each $\tau$, we take $\lambda_{cut}$ with the maximum $\langle\Xi_{AFM}(\tau,\tau')\rangle_{max}$ as the optimal value ($\lambda_{opt}$). At larger flow times it is given by the lowest considered value of $\lambda_{cut}=0.01$.
The maximum $\langle\Xi_{AFM}(\tau,\tau')\rangle_{max}$ starts at a rather small value of 0.4 at $\tau=0$, but increases significantly already at small flow times until it saturates at $\tau\sim3$. This is shown in Fig.~\ref{GF_AFM}, where we also plotted $\langle\Xi_{GF}(\tau,\tau')\rangle$, the correlation between the gauge definition at different flow times $q_c(x,\tau)$ and at the reference flow time $q_c(x,\tau')$. As expected the AFM and the GF converge to the same results at larger $\tau$ and the filtering property of the AFM leads to better results at smaller flow times.

From these results we deduce that for the GF applied before the AFM a flow time of $\tau=0.5$ might be a good compromise: the smearing radius is still small (two lattice spacings), but the eigenvalue spectrum is sufficiently improved to find a reasonable $\lambda_{cut}$ for all configurations.

Note that in the rare cases of configurations with zero content of fractional instantons the correlation $\Xi_{AFM}(\tau,\tau')$ is not meaningful as (\ref{Xi:def}) diverges due to $q_{AFM}$ being completely flat. We had to discard these empty configurations which were chosen according to $|q_c(t,\tau')|<0.1$ for all $t$. For this ensemble 27 were discarded. The AFM at $\tau=0.5$ has for 21 of these configurations no signal, meaning no eigenvalue below the optimal cut. Notice that for physically interesting regimes with high density of fractional instantons these cases are actually marginal.

\begin{figure}[h!]
\centering
\begin{subfigure}{0.45\linewidth}
    \includegraphics[width=\linewidth]{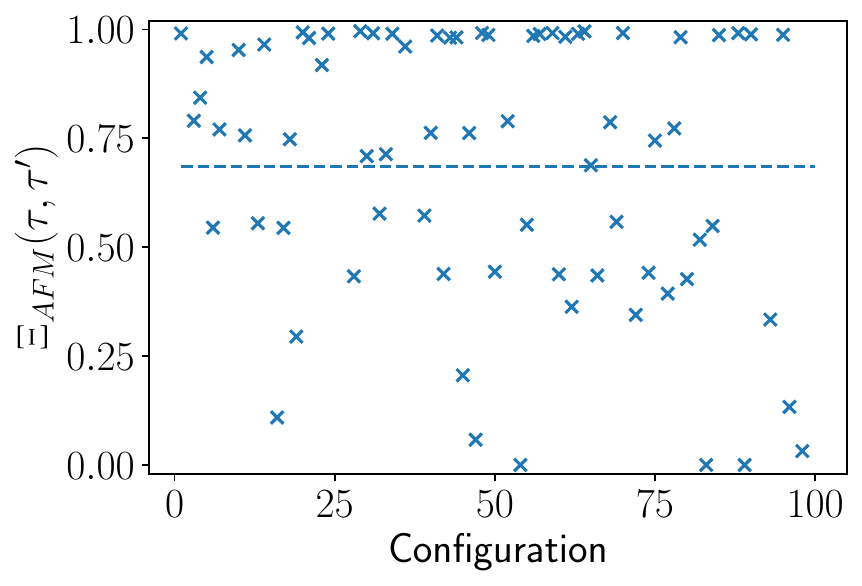}
    \caption{$\tau=0.5$}
\end{subfigure}
\begin{subfigure}{0.45\linewidth}
    \includegraphics[width=\linewidth]{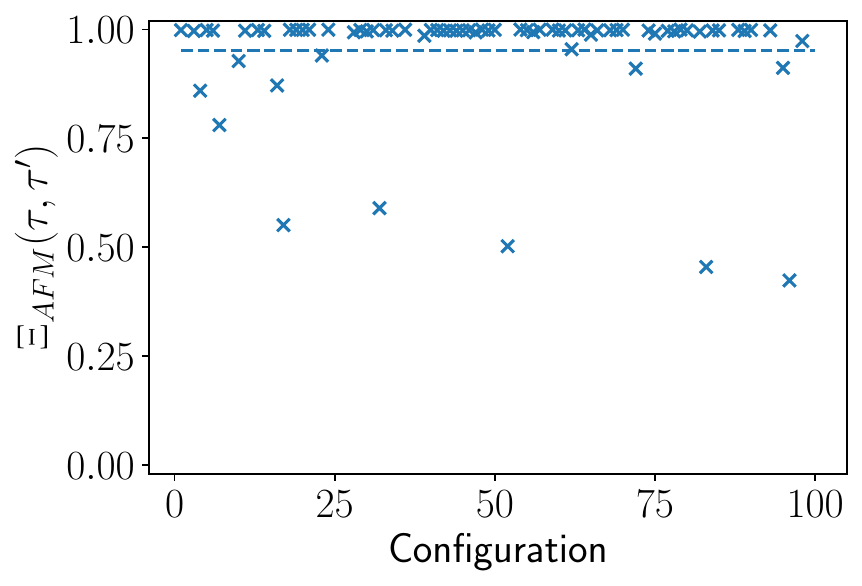}
    \caption{$\tau=2$}
\end{subfigure}

\caption{Monte Carlo history of $\Xi_{AFM}(\tau,\tau')$ between the density obtained from the AFM at $\tau=0.5, 2.0$ and the gauge definition at $\tau'=4$ for the $V=4^3\times 32$ lattice. Dashed lines show the mean values.}\label{GM_history}
\end{figure}

The results obtained so far provide an indication of the viability of the AFM from the average correlation to the GF at large flow times. For a more detailed analysis it is necessary to take a closer look at  individual configurations. Fig.~\ref{GM_history} shows the distribution of $\Xi_{AFM}(\tau,\tau',\lambda_{cut})$ at $\tau=0.5$ and $2.0$ for the whole ensemble with the optimal value $\lambda_{cut}=0.03$ and $0.1$ respectively. It is remarkable that many configurations present a value very close to one $\Xi_{AFM}(\tau,\tau')\sim 1$, but there is also a considerable fraction of configurations with $\Xi_{AFM}(\tau,\tau')<0.6$.

Our quantification of the AFM performance by comparison with the GF does not take into account possible failures of the GF method. We therefore have to inspect the topological charge distributions more closely. 
Fig.~\ref{MC_density} presents some examples of the density profiles $q_{AFM}(x,\tau=0.5)$ and $q_c(x,\tau)$ for $\tau=0.5,2,4$ integrated over spacial coordinates. Several different scenarios can be identified:
\begin{figure}[h!]
\begin{subfigure}{\textwidth}
\includegraphics[width=0.90\textwidth]{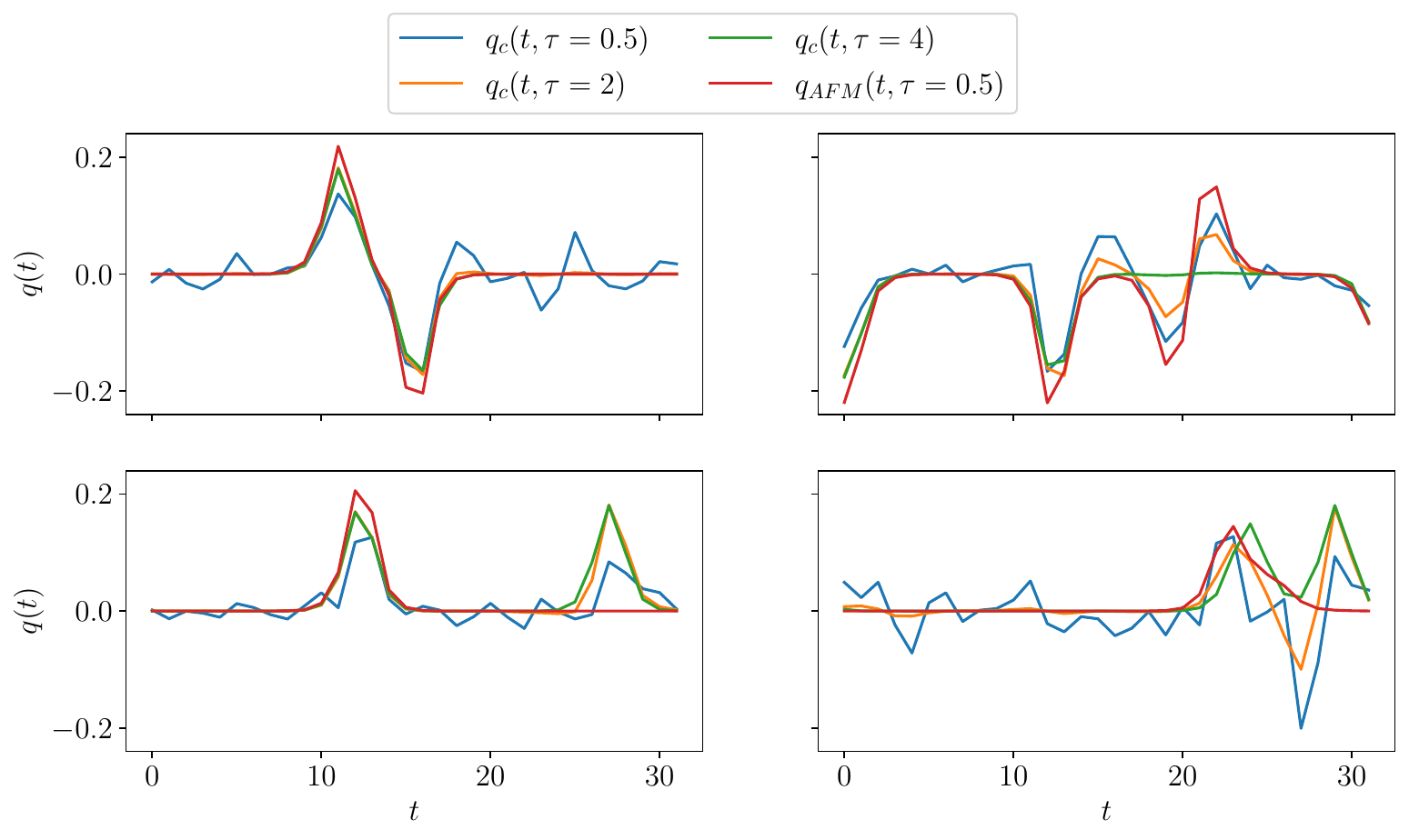}    
\caption{Topological charge densities}
\label{MC_density}
\end{subfigure}
\begin{subfigure}{\textwidth}
\includegraphics[width=0.97\linewidth]{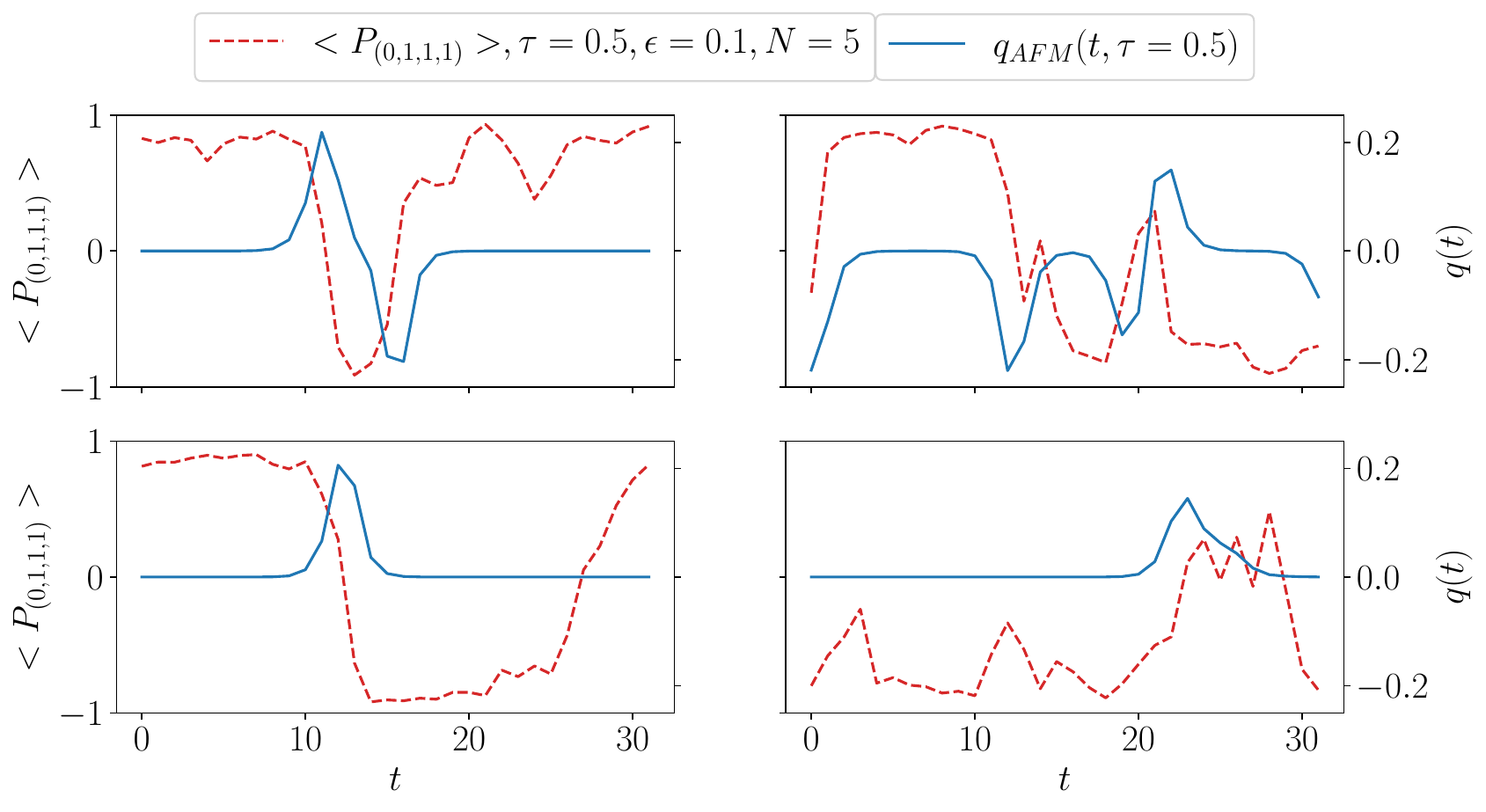}
\caption{Polyakov loop}
\label{Polyakov_Loop}
\end{subfigure}
\caption{(a) Example configurations comparing the topological density integrated over spacial coordinates obtained from the AFM at $\tau=0.5$ with the topological density from the gauge field obtained at different flow times. From left to right, top to bottom, the values of $\Xi_{AFM}=0.99,0.61,0.43,0.24$. (b) Polyakov loop $P_{(0,1,1,1)}$ measured on the same configurations.}
\label{MC_density_all}
\end{figure}
\begin{enumerate}
    \item \textbf{Top-left of Fig.~\ref{MC_density}}: The AFM and the GF both capture distributions of instanton/anti-instantons.
    \item \textbf{Top-right of Fig.~\ref{MC_density}}: Due to instanton/anti-instanton pair annihilation, the GF at large $\tau$ does not capture all structures. The AFM captures all structures.
    \item \textbf{Bottom-left of Fig.~\ref{MC_density}}: The AFM misses structures.\label{point:AFMproblem}
    \item \textbf{Bottom-right of Fig.~\ref{MC_density}}: Both the AFM and the GF seem to have an inconsistent picture and it is difficult to identify the correct topological density.
\end{enumerate}
These observations are also confirmed by the Polyakov loop winding in all spacial directions ($P_{(0,1,1,1)}$). It is expected to vanish at the center of each fractional instanton or anti-instanton and provides an independent signal, see Fig.~\ref{Polyakov_Loop}.

We have seen that the AFM performs reasonably well in many cases, but the significant fraction of the configurations where some sctructures are missed by the AFM, scenario \ref{point:AFMproblem}, needs further discussions. In some of these cases, we were able to identify the missing structures in the densities of higher modes or varying the value of $\kappa$. However, this does not provide a reasonable solution since ``wrong'' contributions are added at the same time by varying these parameters. The problem is hence related to a missing separation between signal and noise contributions.

\begin{figure}[h!]
\centering
\hspace{2cm}
\includegraphics[width=0.75\linewidth]{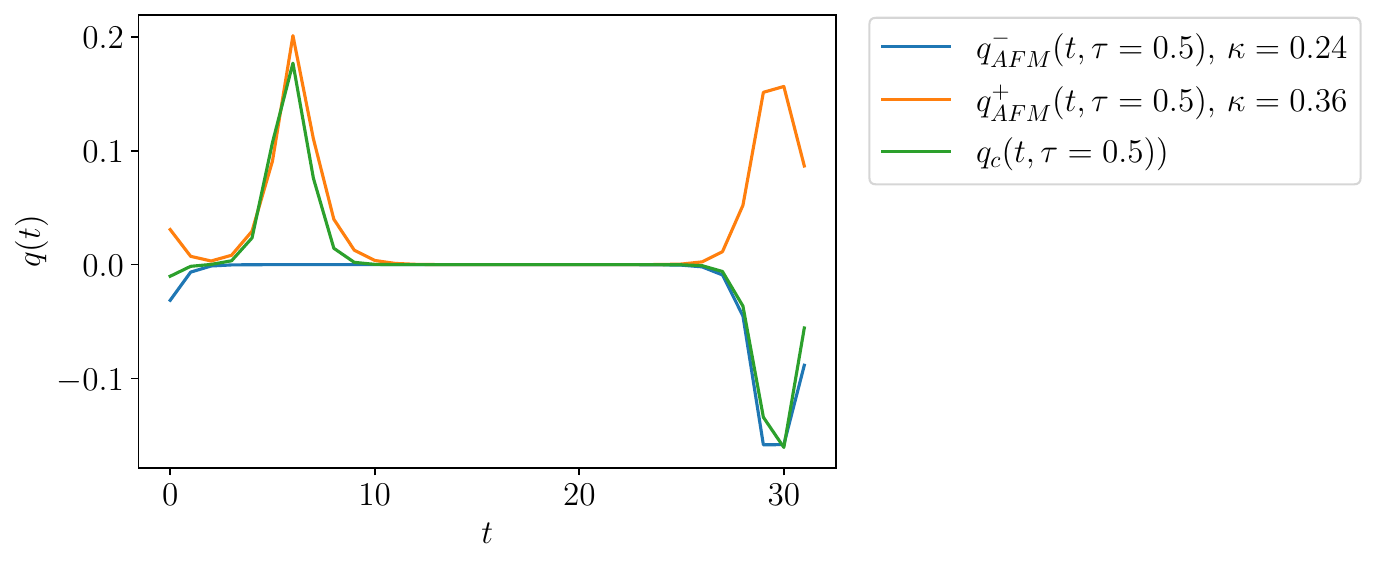}
\caption{Spatial integrated topological charge density from $q_c({t,\tau})$  and $q^{\pm}_{AFM}(t,\tau)$ for $\tau=0.5$ to show the presence of doublers.}
\label{doublers_density}
\end{figure}
The wrong contributions can be in some cases related to the doublers in the Wilson-Dirac spectrum and effects of a tuning of the $\kappa$ parameter. As an example, we computed the lowest eigenmodes of $\Oopm$ at $\kappa_1=0.24$ and $\kappa_2=0.36$ on a configuration with a fractional instanton and a fractional anti-instanton,  Fig.~\ref{doublers_density}. This is reproduced both by the AFM at $\kappa_1$ and the GF density at large flow time. $\kappa_2$ introduces additional zero modes of the overlap operator. The AFM topological charge density in the positive chirality sector, $q_{AFM}^{+}(t,\tau=0.5)$, contains an extra fractional instanton signal. Furthermore the extra instanton has the same position and shape as the fractional anti-instanton. The signal is hence cancelled by the additional wrong contribution. This shows that the densities related to the doublers seem to resemble with high accuracy the density of the physical modes.

\begin{figure}[h!]
\centering
\begin{subfigure}{0.45\linewidth}
    \includegraphics[width=\linewidth]{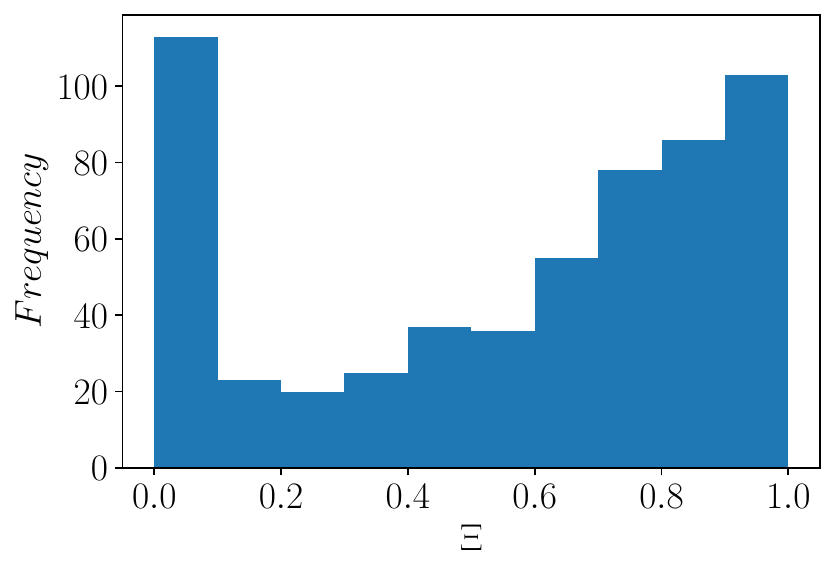}
    \caption{All modes included}
\end{subfigure}
\begin{subfigure}{0.45\linewidth}
    \includegraphics[width=\linewidth]{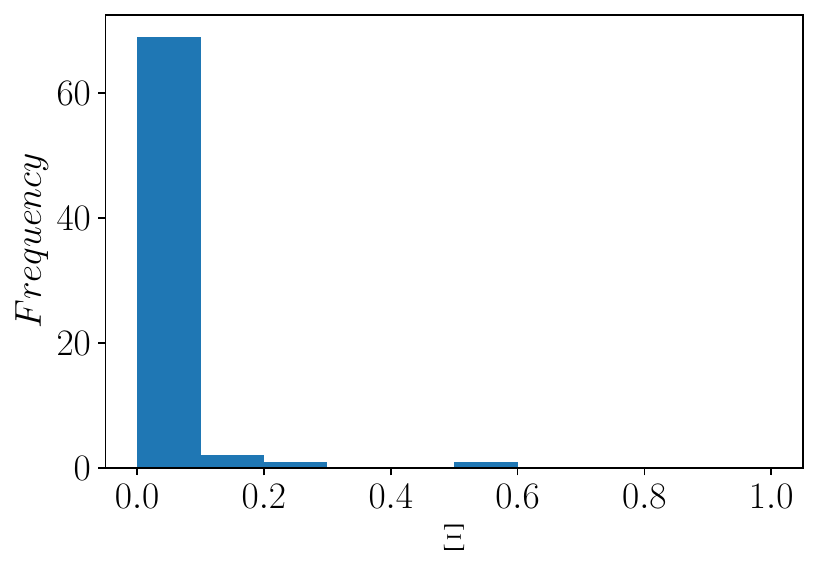}
    \caption{Modes under optimal cut}
\end{subfigure}
\hspace{0.5cm}

\caption{Histogram of cross-correlations $\Xi$ between each left handed and each right handed mode density ($\Oop$ and $\Oom$ operator at $\tau=0.5$) for each configurations. (a) including all 8 modes computed; (b) including only the ones below the optimal cut $\lambda_{opt}$.}
\label{doublers_xi}
\end{figure}
We can push this investigations a bit further and check how much the AFM is affected by the doublers (now for $\kappa=\kappa_1$ and the compete ensemble). If the doublers lead to very similar signals in the opposite chirality sector, they can be identified by the cross-correlations of these sectors, see Fig. \ref{doublers_xi}. If we consider the complete number of eigenmodes (8 per configuration) there is a considerable correlation between the modes on both chiralities. Below our chosen value of $\lambda_{cut}$, this correlations is, however, negligible. This indicates that our optimization of the cut leads to a removal of the doubling modes. At the same time it also removes relevant contributions from the signal. 
From these findings one can already conclude that the AFM results might improve towards the continuum limit. At finer lattices there is a better separation of typical fractional instanton size and lattice spacing. This also means a better separation of physical and unphysical contributions.
\subsection{Results on a finer lattice}
\begin{figure}[h!]
\centering
\begin{subfigure}{0.44\linewidth}
    \includegraphics[width=\linewidth]{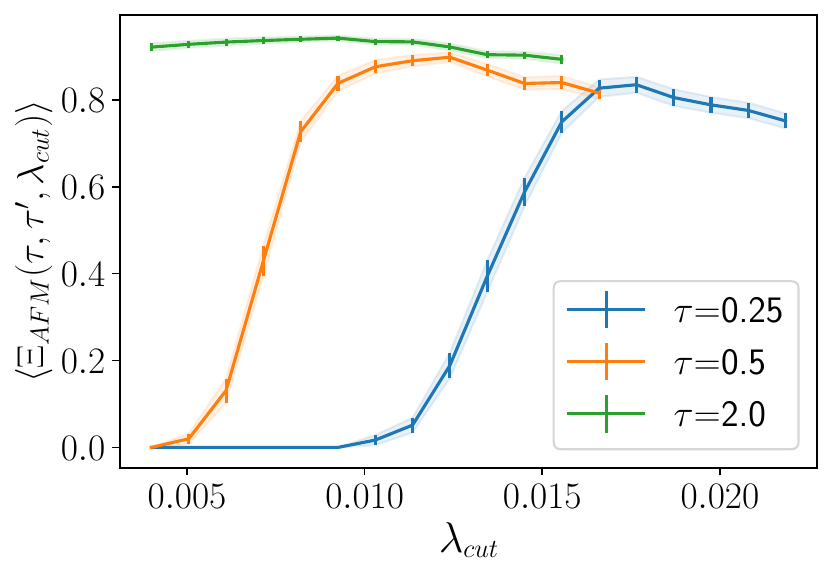}
    \caption{$\tau'=4$}
\end{subfigure}
\begin{subfigure}{0.44\linewidth}
    \includegraphics[width=\linewidth]{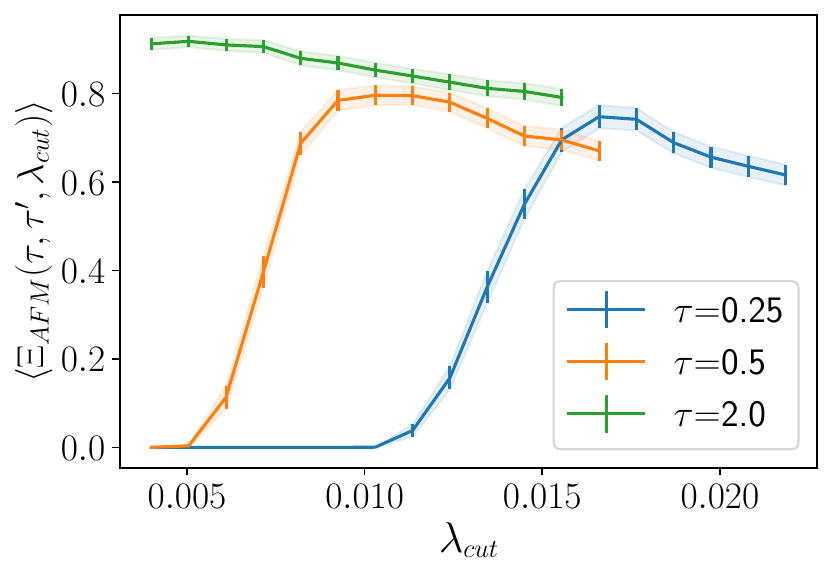}
    \caption{$\tau'=16$}
\end{subfigure}
\caption{Average correlation $\langle\Xi_{AFM}(\tau,\tau',\lambda_{cut})\rangle$ like in Fig.~\ref{Xi} for the finer lattice  ($V=8^3 \times 64$) and two different large reference flow times $\tau'=4,16$.}
\label{Xi_8x64}
\end{figure}

\begin{figure}[h!]
\centering
\includegraphics[width=0.55\linewidth]{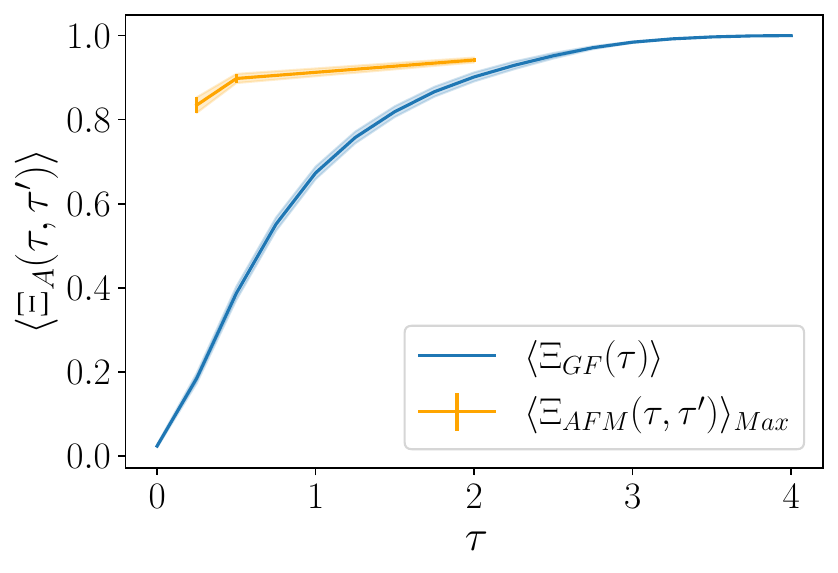}
\caption{Maximum $\langle\Xi_{AFM}(\tau,\tau')\rangle_{max}$ with respect to the cut $\lambda_{cut}$ compared to $\langle\Xi_{GF}(\tau,\tau')\rangle$ for the finer lattice.}
\label{GF_AFM_8x64}
\end{figure}

In order to check the improvements at a finer lattice spacing, we have investigated a lattice $V=8^3\times 64$ at $\beta=2.60$. This means a similar physical volume at half the lattice spacing of previous subsection. At these larger lattices we have not performed the same in-depth analysis due to the additional numerical costs, but it is sufficient for a reasonable comparison. We computed 12 eigenvalues of the $\mathcal{O}^\pm$ for each configuration and we had to discard 21 configurations as they appeared empty. As a reference scale, we have considered $\tau'=4$ and $16$ but the results do not depend much on this choice. The maximum values of $\langle\Xi_{AFM}(\tau,\tau')\rangle_{max}$ in Fig.~\ref{Xi_8x64} have significantly increased with respect to the coarser lattice. Furthremore, the comparison to the GF Fig.~\ref{GF_AFM_8x64} also shows a great increase in performance of the AFM. 

Two different strategies for the scaling of $\tau$ could be applied: in the spirit of an improved lattice action, one would keep the smearing radius in lattice units constant. On the other hand, one could also keep a physical smearing radius constant, which would imply that $\tau$ is rescaled by approximately a factor four compared to the coarse lattice. In both cases much higher values for the correlation are achieved.

\begin{figure}[h!]
\begin{subfigure}{0.49\linewidth}
\includegraphics[width=\linewidth]{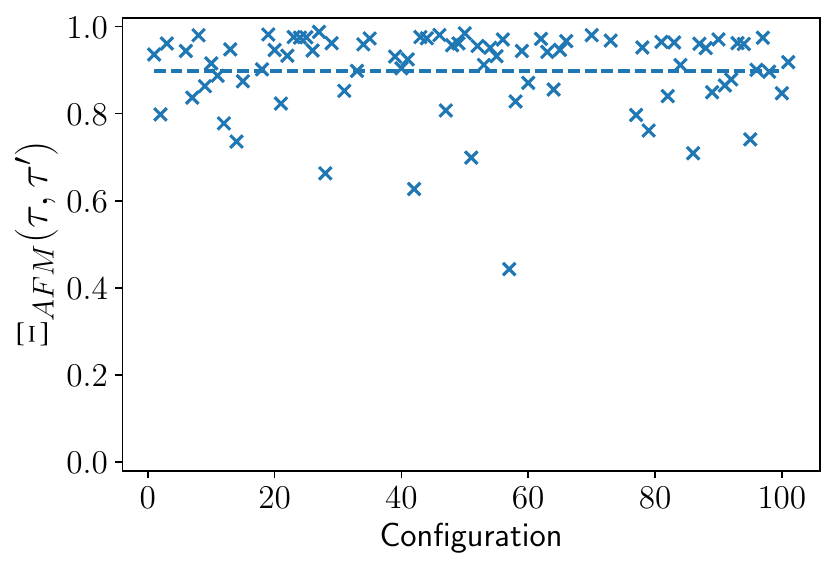}
\caption{$\tau=0.5$}  
\end{subfigure}
\begin{subfigure}{0.49\linewidth}
\includegraphics[width=\linewidth]{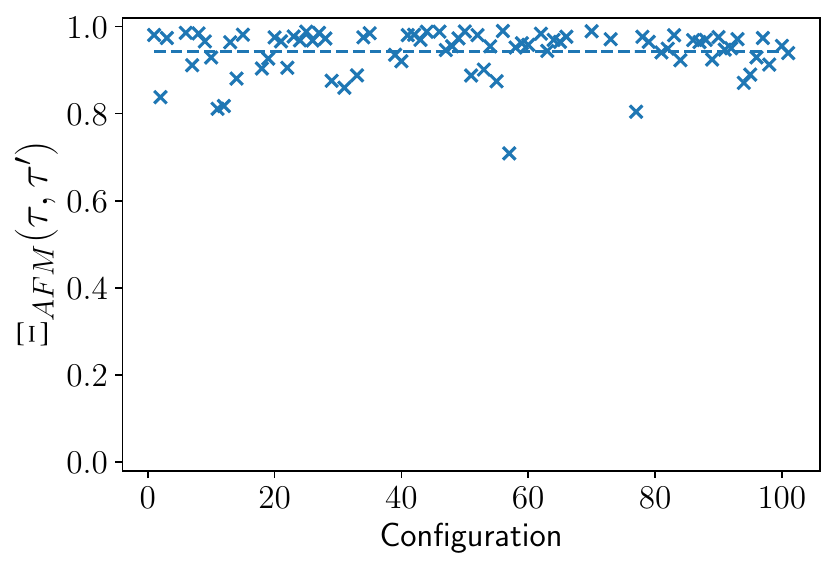}
\caption{$\tau=2.0$}  
\end{subfigure}

\caption{Monte Carlo history of $\Xi_{AFM}(\tau,\tau')$ like in Fig.~\ref{GM_history} for the finer lattice ($V=8^3 \times 64$) and two different flow times $\tau=0.5, 2.0$. The reference flow time is $\tau'=4$.}\label{GM_history_8x64}
\end{figure}

The  improvement is also reflected in the Monte Carlo history of $\Xi_{AFM}(\tau,\tau')$,  Fig~\ref{GM_history_8x64}.  Not only the average value is larger, but also the distribution becomes much narrower. This is a clear indication for a better performance of the AFM at finer lattices. 

\begin{figure}[h!]
\centering
\includegraphics[width=0.90\linewidth]{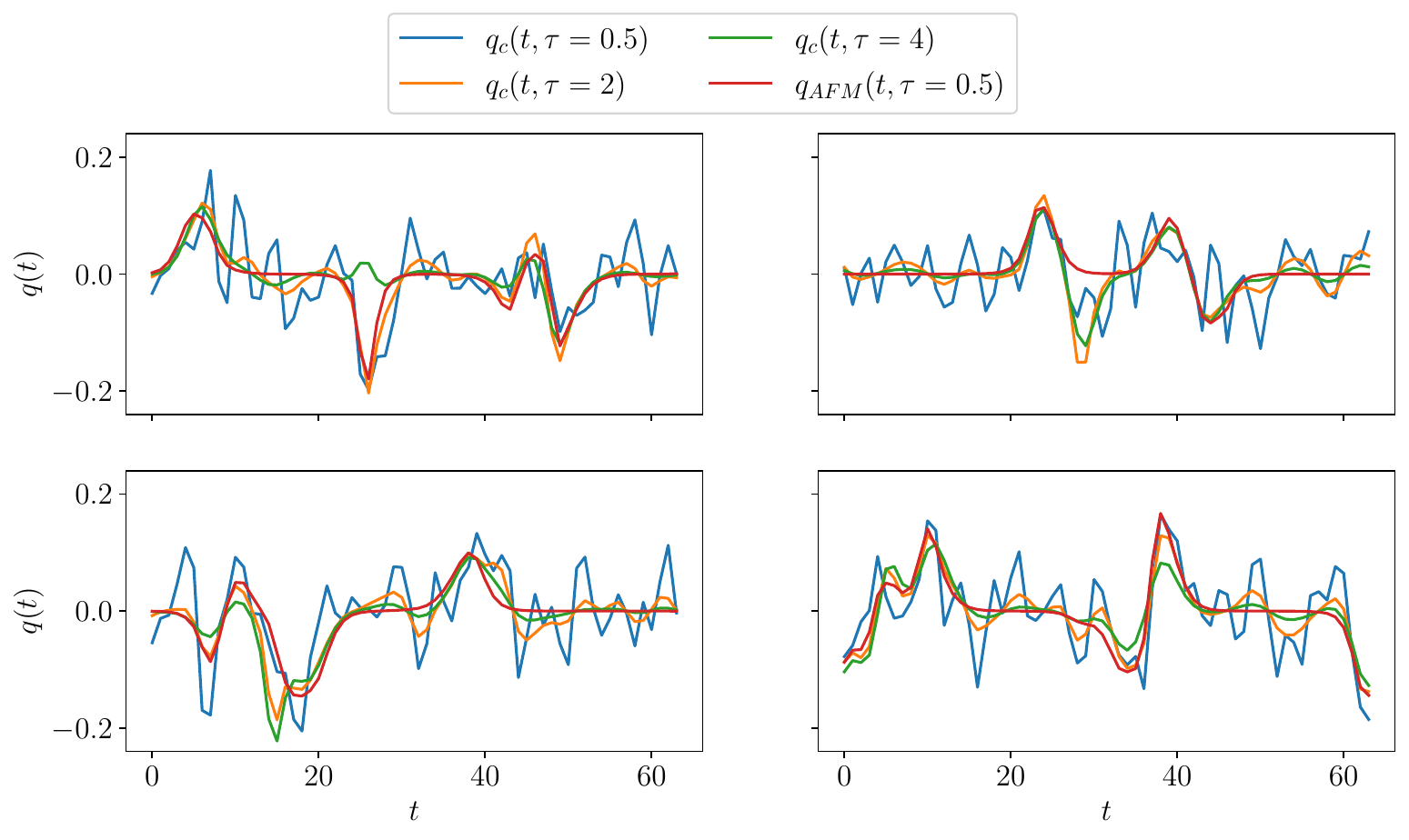}
\caption{Example configurations comparing the spatial integrated topological density obtained from the AFM at $\tau=0.5$ with the topological density from the gauge field obtained at different flow times for the $V=8^3\times 64$ lattice. From left to right, top to bottom, the values of $\Xi_{AFM}(0.5,4.0)=0.44,0.66,0.69,0.85$ (first three values are the smallest of the ensemble).
}
\label{MC_density_8x64}
\end{figure}

Finally, to confirm these findings some configurations are shown in Fig.~\ref{MC_density_8x64}. The sampling of topological structures provides relevant contributions with several structures. The UV fluctuations at small flow times are at much higher momenta due to the smaller lattice spacing.
Even in these examples, the AFM provides a good signal of the distributions and only a small number of structures is missing.

\section{Conclusions and outlook}
The filtering of semiclassical structures can be done using the filtering properties of the lowest modes of the adjoint Dirac operator. The overlap formulation
provides a clean way to project on the chirality sectors and the reality condition optimizes the linear combination of zero modes for smooth configurations. 
We have confirmed in numerical studies of specific smooth configurations resembling fractional instantons that the filtering properties are reproduced on the lattice.
Noise is effectively filtered from the underlying semiclassical content, which we have shown explicitly by adding noise to a smooth configuration. We have observed that after the addition of noise, the signal has to be recovered from several lowest eigenmodes instead of a single SZM.
We have shown that a simple summation of the lowest modes is sufficient to reproduce the topological density. 

Based on these very promising observations, we have considered an application on Monte Carlo generated configurations. At first we have considered a challenging case of a coarse lattice, which allowed to perform an in-depth analysis of the issues of the method. An initial smearing of the configurations at a small smearing radius leads to significant improvements of the method. At large flow time, the method provides the same results as the field theoretical definition of the topological charge density. At small flow time, the method performs very well on a large fraction of the configurations, but there are still a significant number of configurations where it misses relevant parts of the structures.
The most likely explanation for these problems is the missing separation between the scales of the filtered structures and UV noise. Therefore the tuning of $\kappa$ and the cut of the sum of lowest eigenmodes is challenging. 
At a smaller lattice spacing the results are drastically improved. In most cases underlying semiclassical structures can be clearly identified. We were also able to show that the AFM tends to reproduce structures like instanton/anti-instanton pairs significantly better than the GF, which tends to induce annihilations.

As a next step we plan to explore how the method performs in other physical situations, for example at larger volumes. We are also going to investigate ways to reduce the costs of the numerical computation, for example reducing the precision of the representation of the overlap operator.

\section*{Acknowledgments}
G. B.\ and I.S.\ are funded by the Deutsche Forschungsgemeinschaft (DFG) under Grant No.~432299911 and 431842497.
A.G-A acknowledges
support by the Spanish Research Agency (Agencia Estatal de Investigaci\'on) through the grant IFT
Centro de Excelencia Severo Ochoa CEX2020-001007-S, funded by MCIN/AEI/10.13039
/501100011033, and by grant PID2021-127526NB-I00, funded by MCIN/AEI/10.13039/
501100011033 and by “ERDF A way of making Europe”.
Part of the computing time for this project has been provided by the compute cluster ARA of the University of Jena.
\bibliographystyle{apsrev}
\bibliography{references}
\end{document}